\documentclass{article}
\usepackage[final,sglblindworkshop]{neurips_2026}
\workshoptitle{World Models for High-Stakes Health (WMHS)}
\usepackage[utf8]{inputenc}
\usepackage[T1]{fontenc}
\usepackage{amsmath,amssymb,amsthm,booktabs,graphicx,microtype,float}
\usepackage{hyperref,url}
\hypersetup{hidelinks,
  pdftitle={Same Predictions, Different Harms: Causal Auditing of Patient World Models},
  pdfauthor={Yicheng Qi; Xiyi Xiong}}
\newtheorem{theorem}{Theorem}
\newtheorem{proposition}[theorem]{Proposition}

\newcommand{\one}{\mathbf{1}}
\newcommand{\R}{\mathbb{R}}
\newcommand{\E}{\mathbb{E}}
\newcommand{\TV}{\operatorname{TV}}

\title{Same Predictions, Different Harms:\\Causal Auditing of Patient World Models}
\author{%
Yicheng Qi\textsuperscript{1}, Xiyi Xiong\textsuperscript{1}\\
\normalfont\textsuperscript{1}Imperial College London, London, United Kingdom\\
\normalfont yq225@ic.ac.uk, xx725@ic.ac.uk
}
\begin{document}
\maketitle
\begin{abstract}
Patient world models used for clinical trial simulation can agree on transition kernels and arm-specific risks, yet disagree on the fraction of patients harmed by switching treatment---the counterfactual quantity that matters for intervention-aware reasoning. We audit this reliability gap in a two-stage shared-response SCM: a categorical intermediate health state is followed by common terminal care. Under independent stages, the sharp harm interval has closed-form endpoints for at most three intermediate states, with an exactness boundary at four states. Declared dependence and response-mismatch budgets yield calibrated outer bounds when stage independence or complete mediation is relaxed; in a symmetric three-state model the entire sensitivity frontier is sharp, $[0,\min\{1/2,1/3+(\rho+\delta)/2\}]$, and shows exactly how budgets erase the gain over endpoint-only bounds. Two eight-variable response LPs propagate interventional uncertainty for finite-sample audits. Exact witnesses verify attainability. On public clinical simulators (EpiCare; sepsis), native configurations show little resolved stage dependence and no additional joint-compatibility gain over pairwise transport---honest negative results for reliability claims. All experiments are locally reproducible; guarantees remain conditional on the stated causal model. The results provide a concrete protocol for deciding when a patient world model is safe to trust for counterfactual harm.
\end{abstract}

\section{Introduction}
Clinical trial simulation and digital-twin workflows increasingly rely on patient world models that generate longitudinal outcomes under interventions. A common validation practice is to check that two simulators, or a simulator and data, agree on transition frequencies and treatment-specific risks. We argue this is insufficient for high-stakes decisions: two models with identical predictions can still disagree on patient-level counterfactual harm---the fraction of patients with an adverse outcome under regimen $A$ but not $B$. This classical probability of necessity and sufficiency (PNS) depends on how potential outcomes coexist for the same patient. A simulator's shared random seeds impose causal assumptions beyond its transition kernels.

For example, three equiprobable intermediate states and adverse-outcome risk $1/2$ at every state allow independent shared-response mechanisms with harm $H=0$ or $H=1/3$, despite identical arm-specific joint state/outcome tables. This is not a modeling curiosity; it is the reliability gap we audit. Theorems~\ref{thm:three} and~\ref{thm:surface} characterize this ambiguity and its sensitivity to causal assumptions.

We consider an initial treatment, a categorical intermediate health state, and common subsequent care. A terminal response must be reused across every state the two counterfactual trajectories could reach. Independently maximizing each pairwise comparison may therefore describe no single response mechanism. This compatibility problem and mediator-informed causal bounds are established \citep{pmlr-v202-haugh23a,pmlr-v162-gresele22a,Dawid_2022,ijcai2022p376}.

Persistent patient heterogeneity can couple the stages, and the intermediate state may not fully mediate the initial treatment's effect. Two declared budgets make these objections quantitative, bounding stage dependence and treatment-specific response mismatch. Calibrated outer bounds and an attained sensitivity frontier show how the three-state gain can disappear. The budgets remain generally unidentified from observational data.

Our scope is explicit: exact three-state bounds at arbitrary risks, general outer sensitivity bounds with one exact symmetric frontier, and a sharp restricted multistage extension. A small-state confidence projection and two public simulator audits show when additional causal information tightens the bounds and when its value is limited. These guarantees address reliability of intervention-aware patient simulators, not clinical deployment of a treatment policy. Table~\ref{tab:scope} collects the assumptions and guarantees.
\section{Model, information and the sharp target}
Let $\alpha,\beta\in\R^n$ be known probability vectors for $S^A,S^B\in[n]$ under two initial treatments. The common terminal care rule has known adverse-outcome risks $p_i=\Pr(Y=1\mid\operatorname{do}(S=i))$. These primitives require randomized or otherwise causally identified transitions with adequate support. Replacing $p_i$ by observed state-conditionals additionally needs mediator--outcome exchangeability after included baseline covariates; initial-treatment randomization alone is insufficient. Observational conditionals in a confounded or non-Markov representation need not identify the primitives.

The first-stage mechanism is any coupling $C\in\Pi(\alpha,\beta)$. The terminal mechanism is a distribution $Q$ of the response vector $Z=(f(1,U_2),\ldots,f(n,U_2))\in\{0,1\}^n$, constrained by $\E_Q Z_i=p_i$. Independence of the stage-specific exogenous variables requires the joint law $C\otimes Q$. Every such pair is realizable: draw $(S^A,S^B)\sim C$, independently draw $Z\sim Q$, and use the lookup rule $Y^a=Z_{S^a}$. In particular, if both regimes reach the same state, their terminal outcomes coincide.

The harm probability and sharp endpoints are
\begin{equation}
 H(C,Q)=\sum_{i,j}C_{ij}\E_Q[Z_i(1-Z_j)],\qquad
 L=\min_{C,Q}H(C,Q),\quad U=\max_{C,Q}H(C,Q).\label{eq:target}
\end{equation}
With $Y=1$ adverse, marginal risks $m_A=\alpha^Tp$ and $m_B=\beta^Tp$ are identified, but $H$ generally is not. The feasible polytopes are compact and connected; continuity of $H$ ensures its image is the interval $[L,U]$. An upper endpoint is a worst compatible probability, not evidence that this much harm actually occurs.

A pairwise relaxation replaces the single $Q$ by a separately chosen binary coupling for each $(i,j)$. For $i\ne j$, its upper cost is $h^+_{ij}=\min(p_i,1-p_j)$, while consistency requires $h^+_{ii}=0$. The pairwise upper endpoint is
\begin{equation}
 U_{\rm pair}=\max_{C\in\Pi(\alpha,\beta)}\sum_{i,j}C_{ij}h^+_{ij}.\label{eq:pair}
\end{equation}
It is a valid outer bound. We do not attribute a false sharpness claim to methods that already enforce joint response compatibility or condition on a different factual query.

\section{Sharp heterogeneous-risk bounds with three states}
\subsection{The lower endpoint is transport for every state count}
Define $\nu_\alpha=\sum_i\alpha_i\delta_{p_i}$ and $\nu_\beta=\sum_i\beta_i\delta_{p_i}$. Let $W_1$ be one-dimensional Wasserstein distance with absolute-distance cost.
\begin{theorem}[Sharp lower endpoint]\label{thm:lower}
For every $n$ and $p\in[0,1]^n$,
\begin{equation}
 L=\min_{C\in\Pi(\alpha,\beta)}\sum_{i,j}C_{ij}(p_i-p_j)_+
 =\frac{W_1(\nu_\alpha,\nu_\beta)+(\alpha-\beta)^Tp}{2}.\label{eq:lower}
\end{equation}
\end{theorem}
The pairwise lower bound is simultaneously attainable using the shared threshold response $Z_i=\mathbf{1}\{V\leq p_i\}$ for one $V\sim\operatorname{Uniform}(0,1)$ independent of the first stage. The identity $(a-b)_+=(|a-b|+a-b)/2$ gives the second expression. Thus joint compatibility affects the upper and lower endpoints asymmetrically.

\subsection{A cut envelope and its exactness for at most three states}
Write $w=(\alpha+\beta)/2$, and for $S\subseteq[n]$ define $b(S)=\min\{w(S),1-w(S)\}$. The harm identity
\begin{equation}
 H(C,Q)=\tfrac12(\alpha-\beta)^Tp+\tfrac12\Pr_{C\otimes Q}(Y^A\ne Y^B)\label{eq:disagree}
\end{equation}
is useful because a deterministic response vector defines a cut $S=\{i:Z_i=1\}$. The maximum first-stage probability crossing that cut is $2b(S)$. Hence
\begin{equation}
 U\leq U_{\rm cut}:=\tfrac12(\alpha-\beta)^Tp+
 \max_{Q:\E_QZ=p}\sum_{S\subseteq[n]}Q(S)b(S).\label{eq:cut}
\end{equation}
This is a linear program over $2^n$ response patterns. The envelope need not be sharp for general $n$; intersecting it with~\eqref{eq:pair} remains valid.

\begin{theorem}[Three-state sharpness and closed form]\label{thm:three}
For $n\leq3$, $U=U_{\rm cut}$. For $n=3$, put $b_i=\min(w_i,1-w_i)$ and, for distinct $i,j,k$, define
\[
 e_{ij}=\frac{b_i+b_j-b_k}{2},\qquad
 u_{ij}=\min(p_i+p_j,2-p_i-p_j),\qquad
 D=(u_{12}+u_{13}+u_{23}-2)_+.
\]
Then the sharp upper endpoint is
\begin{equation}
 U=\tfrac12(\alpha-\beta)^Tp+
 \sum_{i<j}e_{ij}u_{ij}-\min_{i<j}e_{ij}\,D.\label{eq:three}
\end{equation}
Zero-mass states may be added for $n<3$. Together with~\eqref{eq:lower}, this gives the full sharp interval.
\end{theorem}
\paragraph{Why the envelope is attainable.}
For three states every nontrivial cut is a singleton or its complement. A single transport plan can attain all diagonal lower bounds $C_{ii}=(\alpha_i+\beta_i-1)_+$ simultaneously. It therefore maximizes crossing probability for every cut, so it can be combined with an optimizing response law $Q$ in~\eqref{eq:cut}. No incompatible edge choices are needed.

\paragraph{Why the triangle correction is needed.}
The cut objective equals $\sum_{i<j}e_{ij}\Pr(Z_i\ne Z_j)$, with $e_{ij}\geq0$. Each pair alone permits maximum disagreement $u_{ij}$. A shared triple cannot disagree on all three edges at once; the sum of its edge disagreements is at most two. When the pairwise maxima violate this facet, subtracting the excess $D$ from an edge of minimum weight gives an attainable joint law. The other Bernoulli and triangle constraints remain feasible. Appendix~\ref{sec:supp-cut} supplies the full facet argument and an explicit eight-cell distribution, including boundary cases.

This exactness does not extend unchanged to four states. For
\[
 \alpha=(.1,.1,.4,.4),\quad\beta=(.4,.1,.4,.1),\quad p=(.75,.5,.5,.25),
\]
the exact upper endpoint is $0.30$, while $U_{\rm cut}=U_{\rm pair}=0.35$. A complete rational vertex certificate is supplied in the appendix and code. This identifies the cut reduction's exactness boundary; fixed four-state instances remain computationally manageable.

\subsection{A second exact regime: one dominant pooled state}
The pairwise relaxation is also sharp under a simple concentration condition, without restricting the number of states or terminal risks.
\begin{proposition}[Dominant-state sharpness]\label{prop:dominant}
If $w_k\geq1/2$ for some state $k$, then, for every $p\in[0,1]^n$,
\begin{equation}
 U=U_{\rm pair}=\sum_{i\ne k}\left[\alpha_i\min(p_i,1-p_k)+\beta_i\min(p_k,1-p_i)\right].
 \label{eq:dominant}
\end{equation}
\end{proposition}
For a proof, reroute all non-$k$ transport entries through state $k$. The inequality
$\min(p_i,1-p_j)\leq\min(p_i,1-p_k)+\min(p_k,1-p_j)$ ensures that the pairwise objective cannot decrease, and $\alpha_k+\beta_k\geq1$ supplies enough diagonal mass for the rerouting. The resulting star transport is attained by one shared response: $Z_k=\mathbf1\{V\leq p_k\}$ and $Z_i=\mathbf1\{V>1-p_i\}$ for $i\ne k$, with $V$ independent of the first stage. This sufficient condition is not necessary; Appendix~\ref{sec:supp-dominant} gives the details.

\section{Relaxing the causal assumptions jointly}
\label{sec:sensitivity}
Replace the common response $Z$ by $(Z^A,Z^B)$ with treatment-specific state-response risks $a_i=\E Z_i^A$ and $b_i=\E Z_i^B$. The budgets $\delta_i$ and $\rho$ control response mismatch and stage dependence, respectively. The risks require intervention on both initial treatment and state, or a justified causal identification argument. For $Q=\mathcal L(Z^A,Z^B)$ and $C=\mathcal L(S^A,S^B)$, allow a joint law $J$ satisfying
\begin{equation}
 \Pr_Q(Z_i^A\ne Z_i^B)\leq\delta_i,\qquad
 \TV(J,C\otimes Q)\leq\rho,\qquad \TV(P,R)=\tfrac12\|P-R\|_1.
 \label{eq:budgets}
\end{equation}
Necessarily $\delta_i\geq|a_i-b_i|$. The first budget bounds a direct response change at fixed state; the second bounds dependence between the entire state pair and response tables using a standard TV measure \citep{Bradley2005}. Neither budget is an observed correlation or an identified mismatch probability. Dependence invalidates $m_A=\alpha^Ta$ and $m_B=\beta^Tb$ in general, so we separately retain the regime risks $m_A=\Pr_J(Y^A=1)$ and $m_B=\Pr_J(Y^B=1)$. We assume the specified information is feasible.

Let $U(p)$ denote the independent shared-response upper endpoint at $(\alpha,\beta,p)$. Define
\[
 L_{\rm dir}=\min_{C\in\Pi(\alpha,\beta)}\sum_{i,j}C_{ij}(a_i-b_j)_+,
 \quad \tau_p=\tfrac12\{m_A-m_B-(\alpha-\beta)^Tp\},
\]
\[
 \tau_{\rm dir}=\tfrac12\{m_A-m_B-\alpha^Ta+\beta^Tb\}.
\]
Here $\tau_a$ and $\tau_b$ substitute $p=a$ and $p=b$ into $\tau_p$. Each $\tau$ is half the correction from the corresponding product-model mean contrast to the observed regime contrast; it is determined by these means, not an additional sensitivity budget. The direct-effect lower endpoint $L_{\rm dir}$ is sharp under independence, even with the mismatch budgets: common forward thresholds attain it and have mismatch $|a_i-b_i|$.

\begin{theorem}[Calibrated joint sensitivity]\label{thm:joint}
Every compatible $J$ satisfies
\begin{align}
 H&\geq\max\{0,m_A-m_B,L_{\rm dir}+\tau_{\rm dir}-\rho/2\},\label{eq:jointlo}\\
 H&\leq\min\left\{m_A,1-m_B,
 U(a)+\tau_a+\tfrac12(\rho+\beta^T\delta),
 U(b)+\tau_b+\tfrac12(\rho+\alpha^T\delta)\right\}.\label{eq:jointhi}
\end{align}
A valid upper enclosure may replace either $U$. These bounds are generally outer.
\end{theorem}
\paragraph{Proof idea.}
Under $J$, $H=(m_A-m_B+\Pr_J(Y^A\ne Y^B))/2$. Moving from $J$ to $C\otimes Q$ changes the disagreement event by at most $\rho$. Under the product law, replacing $Z^B$ by $Z^A$ changes it by at most $\sum_j\beta_j\delta_j$; reversing the anchor gives the other inequality. The sharp product lower bound gives~\eqref{eq:jointlo}. Calibrating the mean difference explains the factor $1/2$. The full proof and separate direct/dependence results appear in Appendices~\ref{sec:ext-dependence}--\ref{sec:ext-joint}. Without observed regime means, the shared-response dependence enclosure widens by $\rho$, rather than $\rho/2$, around the original endpoints.

The next result gives an exact robustness law in a symmetric three-state model: each unit of either budget erodes the tightening by $1/2$, until the endpoint-only upper bound $1/2$ is recovered.
\begin{theorem}[Exact joint sensitivity frontier]\label{thm:surface}
For two uniform state laws on three states, $a=b=(1/2,1/2,1/2)$, $m_A=m_B=1/2$, and $\delta_i=\delta$, the entire sharp harm interval is
\begin{equation}
 \left[0,\min\left\{\tfrac12,\tfrac13+\tfrac{\rho+\delta}{2}\right\}\right].\label{eq:surface}
\end{equation}
This remains sharp after fixing both complete arm-specific $(S,Y)$ tables to $\Pr(S=i,Y=y)=1/6$.
\end{theorem}
\paragraph{Attainability.}
Use a uniform cyclic state-pair coupling. Give equal probability to the six nonconstant binary response triples. Initially the response tables agree, yielding harm $1/3$. Mixing in cyclically shifted opposite responses increases harm by $\delta/2$ up to a mismatch of $1/3$. Within the remaining equal-response cells, redistribute probability to disagreeing cells while preserving both marginals and both observed tables. Moving total mass $\rho$ increases harm by $\rho/2$ until saturation at $1/2$. Appendix~\ref{sec:ext-surface} gives explicit nonnegative cells, all intermediate values, and exact rational verification. The bound depends on $\rho+\delta$ in this symmetric example; no universal sum-only reduction is claimed.

\begin{figure}[t]
\centering\includegraphics[width=\linewidth]{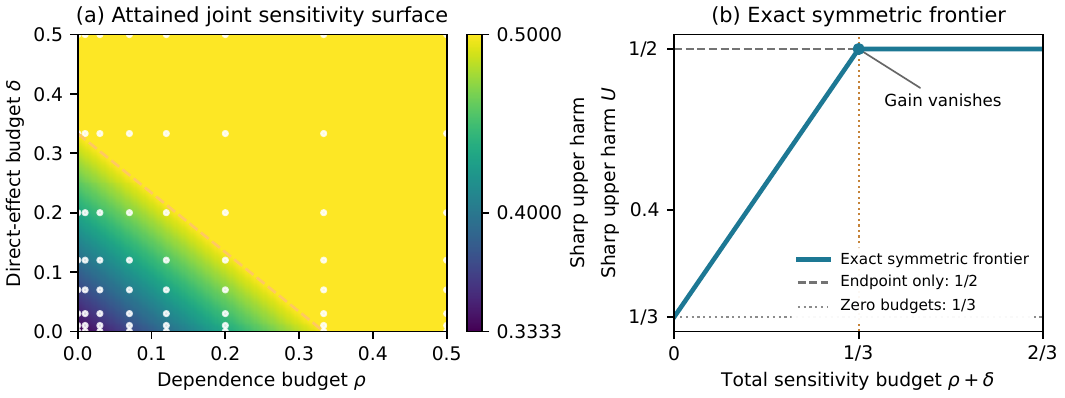}
\caption{Exact sensitivity in the symmetric three-state model of Theorem~\ref{thm:surface}. (a) Sharp upper harm over the two budgets; dots mark 64 independently checked fixed-marginal LPs. (b) The same frontier as a function of total budget: $U$ rises from $1/3$ at slope $1/2$, reaching the endpoint-only upper bound $1/2$ at $\rho+\delta=1/3$. This sum-only law is specific to the stated symmetric model.}
\label{fig:sensitivity}
\end{figure}

\section{Extensions}
\subsection{Multistage continuation}
Suppose the ternary entrance is followed by a binary state $T$, with $k_i=\Pr(T=1\mid\operatorname{do}(S=i))$, and then $h$ binary Markov transitions under common care and time-specific noises independent of each other and of the entrance. Common care may depend on time and current state, but not unrecorded history. Let $[L_T,U_T]$ be the sharp entrance interval, $\Delta_T=(\alpha-\beta)^Tk$, and $D_- =2L_T-\Delta_T$, $D_+=2U_T-\Delta_T$. At suffix step $t$ let the two risks be $(r_{t0},r_{t1})$, and put $a_t=|r_{t1}-r_{t0}|$, $b_t=\min(r_{t0}+r_{t1},2-r_{t0}-r_{t1})$. Then
\begin{equation}
 L_h=\tfrac12\left(\Delta_h+D_-\prod_{t=1}^h a_t\right),\quad
 U_h=\tfrac12\left(\Delta_h+D_+\prod_{t=1}^h b_t\right),\quad
 \Delta_h=\Delta_T\prod_{t=1}^h(r_{t1}-r_{t0}).\label{eq:suffix}
\end{equation}
This connects the ternary entrance to the established binary-chain disagreement product \citep{Dawid_2022}; it does not solve arbitrary multivalued trajectories or relax suffix independence. The proof uses the fact that equal paired states remain equal, whereas unequal binary states disagree next with a probability in $[a_t,b_t]$. Each endpoint is jointly attainable. Known intermediate kernels must be retained: for $\alpha=\beta=(.2,.3,.5)$, $k=(.95,.05,.95)$, and one final transition $(.5,.5)$, the true interval is $[0,.32]$, while collapsing to state-specific final risks gives $[0,.5]$. Information loss from collapsing mediators is already known for binary chains; this is a concrete ternary-entrance instance.

\subsection{Finite-sample uncertainty}\label{sec:finite-sample}
For the independent shared-response model with three states, define $c_i(z)=1$ when state $i$ is the unique one or the unique zero in response triple $z$; all $c_i$ vanish for constant triples. A transport with $C_{ii}=\min(\alpha_i,\beta_i)$ simultaneously minimizes every cut crossing. Consequently, the centered lower endpoint also has an exact response LP:
\[
 L-\tfrac12(\alpha-\beta)^Tp=\min_{Q:\E_QZ=p}\sum_{z,i}Q(z)c_i(z)\tfrac12|\alpha_i-\beta_i|.
\]
Both sharp endpoints thus have a small response-LP representation. This lets us propagate probability boxes without general nonlinear causal optimization or a sum of separate Lipschitz errors. Let simultaneous boxes be denoted $[\alpha_i^-,\alpha_i^+]$, with analogous notation for $\beta,p,m_A,m_B$, and set
\[
 \underline d_i=\tfrac12\max(0,\alpha_i^- -\beta_i^+,\beta_i^- -\alpha_i^+),\quad
 \overline b_i=\min\{\tfrac12,(\alpha_i^++\beta_i^+)/2,1-(\alpha_i^-+\beta_i^-)/2\}.
\]
Over $Q\ge0$ with total mass one and $p_i^-\le\E_QZ_i\le p_i^+$, solve two eight-variable LPs:
\[
 \ell=\min_Q\sum_{z,i}Q(z)c_i(z)\underline d_i,\qquad
 u=\max_Q\sum_{z,i}Q(z)c_i(z)\overline b_i.
\]
Writing $\Delta_-=m_A^--m_B^+$ and $\Delta_+=m_A^+-m_B^-$, the confidence outer interval is
\begin{equation}
 [\ell+\Delta_-/2,u+\Delta_+/2]\cap
 [\max(0,\Delta_-),\min(m_A^+,1-m_B^-)].\label{eq:fullci}
\end{equation}
Every true primitive in the boxes satisfies this enclosure. Compatible point boxes recover both sharp endpoints; nonzero boxes need not give sharp nuisance-robust extrema. We use two-sided Clopper--Pearson intervals with failure $\zeta/(3n+2)$ per coordinate for the $n$ state probabilities in each arm, $n$ state-intervention risks and two endpoint means. A union bound gives simultaneous coverage at least $1-\zeta$ without requiring independence between confidence events. Observations must be iid within their relevant batches. For $n<3$, zero-mass padding applies; an unobserved state intervention receives $[0,1]$. This specializes established confidence-region propagation \citep{Duarte2024} to the small response LPs. Appendix~\ref{sec:ternary-projection-proof} gives the proof and numerical scope. The more general Lipschitz/Hoeffding wrapper, including joint-sensitivity propagation, remains in Appendix~\ref{sec:ext-finite-sample}; the present CP projection assumes the original independent shared-response model.

\section{Local verification and public health audits}
All computations use a local Apple M4 CPU with 24\,GiB RAM, public source code and generated probabilities. The package includes seeds, retained cases, licenses and a single CPU reproduction command.

\paragraph{Auditing protocol.}
The empirical sections follow a fixed protocol for patient world models. (i) Declare the information set: first-stage state laws, interventional terminal risks, and optional regime means. (ii) Compute sharp or outer harm intervals under the independent shared-response model. (iii) Report sensitivity over declared dependence and response-mismatch budgets $(\rho,\delta)$. (iv) Label each result as sharp or outer; retain attaining witnesses when available. (v) Compare against pairwise transport and endpoint-only baselines at equal information or equal observation counts. (vi) Disclose null results and keep a local reproduction record.

\paragraph{Algebra, sharpness and generality.}
The baseline is checked against exhaustive transport/response vertex pairs at up to four states and 4,500 additional one-to-three-state formula checks. The four-state counterexample has an exact rational certificate. For the joint surface, 81 rational grid points verify nonnegativity, all marginals, full observed tables and actual TV; 64 floating-point LPs separately hold $C,Q$ fixed and agree. This fixed-marginal problem is linear, whereas optimizing unknown $C,Q$ inside $\TV(J,C\otimes Q)$ is generally not a single LP. We test 150 joint mechanisms at $n\in\{1,2,3,4,6\}$ with arbitrary risks and dependence. Every realized harm lies in the proposed outer interval.

For 126 arbitrary-risk direct-effect configurations, enumerating all transport vertices and solving a $4^n$-cell response LP at each yields global numerical references. The general analytic upper enclosure exceeds these references by mean $.03271$ and maximum $.19749$ (Appendix Figure~\ref{fig:direct-reference}); it is not generally sharp. Thirty-two multistage cases with suffix lengths $1,2,4,6$ agree with exhaustive vertex composition. Maximum LP primal residual is $6.50\times10^{-15}$; a small residual is a numerical diagnostic, not a formal outward-rounded upper certificate.

\paragraph{Unknown probabilities and equal observation budgets.}
We use six arbitrary-risk settings, including a $0/1$ boundary setting, $N\in\{100,500,2000,10000\}$ and 1,000 repetitions per setting. Each arm supplies $N$ joint state/outcome observations, and each of three separate state interventions supplies $N$ outcomes: total $5N$. All 24,000 CP-projection intervals in~\eqref{eq:fullci} contain the full true identified range. This demonstrates conservatism, not exact 95\% calibration. Mean widths improve over the retained Lipschitz/Hoeffding wrapper in all 24 strata, by $.00457$--$.11633$. Three hundred point-box checks match global vertex extrema within $2.99\times10^{-12}$; 300 general-box checks have no enclosure violation.

At equal $5N$ observation counts, each endpoint-only arm adds $3N/2$ independent outcomes to its original $N$. CP and Hoeffding rectangles and a stronger three-event comparator each guarantee at least 95\% full-range coverage; the latter combines one-sided contrast and exact arm-risk bounds. The CP projection is narrower on average than the stronger three-event comparator in 12 of 24 strata and wider in 12. At $N=100$, endpoint-only inference wins in all six settings; at $N=10000$, the projection wins in five, reducing mean width by $.10287$--$.17841$, while setting 6 remains $.00571$ wider. Equal observation counts do not imply equal clinical intervention costs. Appendix~\ref{sec:budget-experiments} gives the protocol, guarantees, coverage targets and complete results.

\begin{figure}[t]
\centering\includegraphics[width=\linewidth]{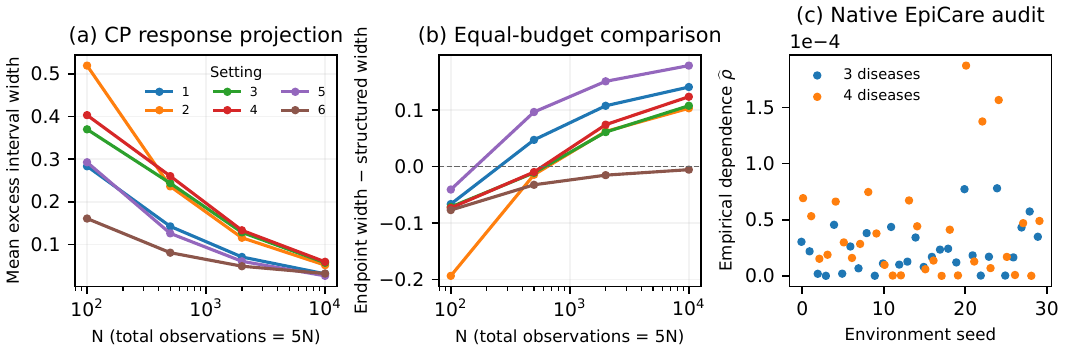}
\caption{(a) CP-projection width above the true identified range. (b) Width of the stronger three-event endpoint comparator minus projection width, at equal $5N$ observation counts; negative values expose the small-sample cost of unknown mediator primitives. (c) Near-zero empirical dependence is the native audit outcome across all 60 persistent-modifier EpiCare cases. Empirical TV is a diagnostic; every full-table pointwise 99\% lower bound is zero. Separate simultaneous primitive bounds support population inference.}
\label{fig:uncertainty-native}
\end{figure}

\paragraph{Native EpiCare: a dependence audit with small effects.}
The two-visit public EpiCare audit treats the decision simulator as a patient world model and retains native adverse-event termination and persistent patient modifiers \citep{NEURIPS2024_eb868633}. Three/four disease states plus an adverse-event absorber give four/five states. First actions have zero immediate remission; the endpoint is an adverse event by visit two, without survival conditioning. Both regimes use the same predeclared state-adaptive terminal care. Thirty seeds crossed with disease count and modifier on/off yield 120 cases. Source-compatible shared ranks specify one explicit counterfactual SCM; the kernels alone do not identify it. Joint count tables retain all $1{,}048{,}576$ simulated patients per case (125.83 million total).

The native audit returns little resolved dependence: with persistent modifiers enabled, mean empirical TV is $2.37\times10^{-5}$ for three diseases and $4.02\times10^{-5}$ for four; the maximum is $1.87\times10^{-4}$. The full-table pointwise 99\% lower dependence bound is zero in every case. These results do not establish a practically large violation. Clopper--Pearson boxes propagate all primitive and dependence-budget uncertainty; the appendix separates population confidence bounds from plug-in intervals. Eight native-source checks (80,000 episodes) agree with the vectorized endpoint frequencies within native 99\% intervals. Four/five-state cut bounds are labeled outer unless an exactness condition applies.

\paragraph{Independent public benchmark: an applicability boundary.}
The public sepsis simulator of \citet{oberst2019counterfactualoffpolicyevaluationgumbelmax}, widely used for counterfactual off-policy evaluation of treatment regimes, retains initial treatment flags in the intermediate state. The compared initial actions therefore have disjoint nonabsorbing state supports. A shared response can use forward thresholds on one support and reverse thresholds on the other, attaining every pairwise upper cost simultaneously; shared absorbing states have deterministic outcomes. Thus this setup has no additional joint-compatibility gain beyond pairwise transport. Sixty predeclared native configurations and 240,000 episodes corroborate support separation using the complete sufficient state. This negative applicability audit concerns gain over pairwise transport; endpoint-only bounds may still be looser.

A separate EpiCare-derived remission stress experiment retains 720 cases and explicitly modifies transition diffuseness. All 240 upstream-transition cases have zero pairwise gap; only declared stress cases give positive gaps, at most $.11347$. The complete protocol and old balanced-risk diagnostic are in Appendices~\ref{sec:legacy-evaluation} and~\ref{sec:experiments-supp}. Native and modified results are reported separately.

\section{Related work}
The target $\Pr(Y^A=1,Y^B=0)$ is the classical probability of necessity and sufficiency (PNS), here oriented toward an adverse outcome and framed as an audit of patient world models for intervention-aware simulation. \citet{Kallus2022Harm} derives sharp bounds on the fraction negatively affected by treatment using observed covariates and develops robust inference with learned nuisance functions. We complement that line by restricting how mediator co-responses couple inside a two-stage simulator, rather than replacing the estimator: explicit three-state endpoints, an attained sensitivity frontier, and a finite-sample set-containment construction without claiming optimal statistical efficiency.
\citet{Dawid_2022} derive sharp bounds for binary complete-mediator chains. \citet{ijcai2022p376} provide graphical bounds for pure mediation, and \citet{shu2026generalprobabilitiescausationcausal} extend mediator-informed probabilities of causation. \citet{xie2026boundingprobabilitiescausationpartial} incorporate partial causal information through optimization and identify nonlinear mediator constraints as a computational difficulty. Our closed-form multivalued shared-response result is complementary to that general optimization program: we solve a health-simulation subclass exactly and mark where the cut reduction fails. The four-state example is a limit of this reduction, not a general computational dichotomy.

\citet{Pena2023} bounds benefit and harm using sensitivity parameters for unmeasured exposure--outcome confounding, while \citet{DingVanderWeele2016} derive sharp sensitivity bounds for mean direct and indirect effects under mediator--outcome confounding. \citet{KawakamiTian2025,pmlr-v286-kawakami25b} define direct and path-specific PNS and establish identification under additional assumptions. We retain the total harm query and bound deviations from stage independence or equality of treatment-specific response tables. The general stability inequalities follow from calibrated outcome means and disagreement probabilities; our attainable joint sensitivity surface is specific to the symmetric half-risk three-state family. The dependence budget uses a standard total-variation dependence measure, not an observable correlation coefficient. Our binary-suffix extension combines the three-state entrance bound with the established product rule for independent binary mediator chains \citep{Dawid_2022}; that work also studies information loss from collapsing mediators.

For dynamic counterfactuals, \citet{oberst2019counterfactualoffpolicyevaluationgumbelmax} specify Gumbel-Max mechanisms, whereas \citet{pmlr-v202-haugh23a} optimize over compatible mechanisms and exhibit incompatibility of pairwise response marginals---the reliability language closest to auditing clinical world models. Structural marginal compatibility is treated more generally by \citet{pmlr-v162-gresele22a}. Our prospective two-regime query differs from the factual-trajectory-conditioned query of \citet{lally2026robustcounterfactualinferencemarkov}. Cut representations of symmetric Bernoulli dependence \citep{Huber_2019} and balanced-subset constructions in experimental design \citep{morrison2025optimalcorrelationbernoullitrials} provide mathematical precedents for our cut and subset-sum formulations. Finally, \citet{Duarte2024} develop general discrete causal optimization and conservative inference by propagating confidence regions for observed probabilities. Following this principle, our specialized endpoints propagate uncertainty in all model primitives without a general nonlinear solver.

\section{Limitations and responsible use}
\paragraph{Using the bound as an assumption audit.}
In the symmetric example, declaring $\rho=\delta=.05$ raises worst compatible harm from $1/3$ to $23/60\approx.3833$, retaining a gain of $7/60\approx.1167$ over the endpoint-only upper bound. The gain disappears at total budget $1/3$. These are conditional statements, not clinical thresholds. An audit should establish its interventional information set, report sensitivity across declared budgets, and retain known intermediate kernels. A simulator-selected coupling remains additional causal structure.

\begin{table}[h]
\centering\caption{Scope of the principal guarantees. ``Sharp'' means an attaining common mechanism exists, conditional on the specified information and model.}\label{tab:scope}
\begin{tabular}{p{.43\linewidth}p{.49\linewidth}}
\toprule
Model class & Guarantee and remaining restriction\\
\midrule
Independent shared response, $n\le3$ & Sharp endpoints; arbitrary risks.\\
Independent shared response, general $n$ & Sharp lower; cut/pair upper generally outer. Exact under a dominant pooled state.\\
Joint dependence and direct-response budgets & General calibrated outer interval; sharp symmetric three-state frontier.\\
Ternary entrance and binary Markov suffix & Sharp at any suffix length; requires independent noises and common care.\\
Unknown interventional primitives & Conservative simultaneous confidence outer interval under valid sampling.\\
\bottomrule
\end{tabular}
\end{table}

General sensitivity bounds can be loose; the sharp joint frontier is symmetric. Valid interventional primitives, sufficient states and declared budgets remain necessary. Native audits provide limited evidence of material gains, and confidence intervals can be wide. These are conditional auditing guarantees for patient world models, not a validated digital twin or a complete treatment-selection rule.

\subsection{Responsible use}
We use generated probabilities and public simulator code, without identifiable patient records. Outputs must not support dosing, clinical deployment or patient-benefit claims. Misstated mediation, independence or sensitivity assumptions can invalidate safety conclusions; explicit information sets, attaining witnesses, outer-bound labels and reproduction records help expose this risk. AI tools assisted research, proofs, code and writing; accountable human authorship and independent scientific review remain necessary. Experiments ran locally; the AI assistant is not claimed to run offline.
\label{mainend}
\clearpage
\bibliographystyle{plainnat}
\bibliography{references}
\clearpage
\appendix
\raggedbottom
\section{Model, realizability, and the identified interval}
\label{sec:supp-model}
Write $[n]=\{1,\ldots,n\}$ and let $\alpha,\beta$ be probability vectors on $[n]$.
Let $\Pi(\alpha,\beta)$ denote the transportation polytope with these row and column marginals, and let
\[
 \mathcal Q(p)=\left\{Q\in\mathcal P(\{0,1\}^n):\ \mathbb E_Q Z_i=p_i\text{ for all }i\right\}.
\]
The admissible mechanisms are pairs $(C,Q)\in\Pi(\alpha,\beta)\times\mathcal Q(p)$.
The first-stage potential states have law $C$, while the response vector $Z$ has law $Q$ and is independent of the first-stage potential states.
The shared terminal assignment is $Y^A=Z_{S^A}$ and $Y^B=Z_{S^B}$.
Thus
\begin{equation}
 H(C,Q)=\sum_{i,j}C_{ij}\mathbb E_Q[Z_i(1-Z_j)].
 \label{eq:supp-objective}
\end{equation}
Set $L=\min H(C,Q)$ and $U=\max H(C,Q)$ over the stated product of polytopes.

\paragraph{Realizability.}
Every admissible pair is a structural causal model. Independently sample $U_1=(I,J)\sim C$ and $U_2=Z\sim Q$; define the first-stage structural function to return $I$ under treatment $A$ and $J$ under treatment $B$, and define the shared terminal function by $f(s,U_2)=Z_s$.
Conversely, every structural model with the stipulated independence and shared assignment induces such a pair $(C,Q)$.
Consequently this representation neither discards nor adds mechanisms within the stated model class.
The first-stage and terminal polytopes are nonempty, compact, and convex. Their product is path connected and the objective is continuous. Its image in $\mathbb R$ is therefore the compact interval $[L,U]$.
This is a prospective fully mediated model; conditioning on an individual factual trajectory would change the feasible problem.

\paragraph{A useful identity.}
The terminal marginal risks are $m_A=\alpha^\top p$ and $m_B=\beta^\top p$, independently of the unknown mechanism.
For every compatible pair,
\begin{equation}
 H(C,Q)=\frac12\left\{(\alpha-\beta)^\top p+
       \mathbb P_{C\otimes Q}(Z_{S^A}\ne Z_{S^B})\right\}.
 \label{eq:supp-disagreement}
\end{equation}
This follows by subtracting the two directional disagreement probabilities.

\section{The exact lower endpoint for arbitrary state spaces}
\label{sec:supp-lower}
For every $n$ and $p$,
\begin{equation}
 L=\min_{C\in\Pi(\alpha,\beta)}\sum_{i,j}C_{ij}(p_i-p_j)_+
  =\frac12\left\{W_1\left(\sum_i\alpha_i\delta_{p_i},
                        \sum_j\beta_j\delta_{p_j}\right)
                   +(\alpha-\beta)^\top p\right\}.
 \label{eq:supp-lower}
\end{equation}
\paragraph{Proof.}
The Fr\'echet inequality gives $\mathbb E[Z_i(1-Z_j)]\ge(p_i-p_j)_+$ for every pair.
One shared response construction attains all these inequalities simultaneously: take a uniform random variable $V$, independent of the first-stage noise, and set $Z_i=\mathbf 1\{V\le p_i\}$.
It follows that minimization over $C$ is exact.
The identity $(x-y)_+=(|x-y|+x-y)/2$ then gives the one-dimensional transportation expression.
Sorting the risk values and transporting mass in increasing order evaluates it in $O(n\log n)$ time.\hfill$\square$

\section{Cut transport and the three-state formula}
\label{sec:supp-cut}
Set $w=(\alpha+\beta)/2$ and
\[
 b_w(S)=\min\{w(S),1-w(S)\},\qquad
 b(w)=\max_{S\subseteq[n]}b_w(S).
\]
For a binary assignment $z$, write $S(z)=\{i:z_i=1\}$.

\paragraph{Lemma: maximal transport across a fixed cut.}
For every $S\subseteq[n]$,
\begin{equation}
 \max_{C\in\Pi(\alpha,\beta)}
   \mathbb P_C\bigl(\mathbf 1\{S^A\in S\}\ne\mathbf 1\{S^B\in S\}\bigr)
   =2b_w(S).
 \label{eq:supp-cut-transport}
\end{equation}
\paragraph{Proof and witness.}
Let $a=\alpha(S)$ and $b=\beta(S)$. Aggregate a coupling into groups $(S,S^c)$.
The maximal off-diagonal mass of the resulting $2\times2$ table is
$\min(a+b,2-a-b)=2b_w(S)$.
It is attained by
\[
 T_{11}=(a+b-1)_+,\quad T_{10}=\min(a,1-b),\quad
 T_{01}=\min(1-a,b),\quad T_{00}=(1-a-b)_+.
\]
Disaggregate a block $(g,h)$ as
$C_{ij}=T_{gh}\alpha_i\beta_j/[\alpha(g)\beta(h)]$ for $i\in g,j\in h$.
A block with zero marginal mass is zero. This preserves both state marginals.\hfill$\square$

\paragraph{The cut-envelope bound.}
For arbitrary $n$,
\begin{equation}
 U\le U_{\rm cut}:=\frac12(\alpha-\beta)^\top p+
       \max_{Q\in\mathcal Q(p)}\mathbb E_Q b_w(S(Z)).
 \label{eq:supp-cut-envelope}
\end{equation}
Indeed, apply~\eqref{eq:supp-cut-transport} to each response assignment in~\eqref{eq:supp-disagreement}.
The bound permits a separate first-stage maximizing coupling for each cut; this is the only relaxation in this step.

\paragraph{Lemma: simultaneous minimum diagonal.}
There exists $C_*\in\Pi(\alpha,\beta)$ with
\begin{equation}
 (C_*)_{ii}=(\alpha_i+\beta_i-1)_+\quad\text{for every }i.
 \label{eq:supp-min-diag}
\end{equation}
\paragraph{Proof.}
At most one index has $\alpha_i+\beta_i>1$.
If none does, consider the bipartite transportation network with diagonal edges forbidden.
For a singleton row $i$, the available column capacity is $1-\beta_i\ge\alpha_i$.
Every row subset containing at least two indices has every column as a neighbor, so it imposes no additional restriction.
The max-flow/min-cut criterion therefore gives a zero-diagonal coupling.
If an index $k$ has excess $d=\alpha_k+\beta_k-1>0$, put mass $d$ on $(k,k)$ and reduce the corresponding row and column masses by $d$.
The remaining total mass is $1-d$; the singleton capacity inequalities still hold, with equality for $k$, so the same argument gives a zero-diagonal residual coupling.
The $n=1$ case is immediate.\hfill$\square$

For $n\le3$, every nontrivial cut is a singleton or its complement.
The coupling in~\eqref{eq:supp-min-diag} maximizes every such crossing probability simultaneously, since the crossing mass at singleton $i$ is $\alpha_i+\beta_i-2(C_*)_{ii}$.
Thus
\begin{equation}
 U=U_{\rm cut}\qquad(n\le3),
 \label{eq:supp-three-exact}
\end{equation}
for every heterogeneous risk vector $p$.

\paragraph{Closed form for three states.}
For $i\in\{1,2,3\}$, put $b_i=\min(w_i,1-w_i)$.
For $i<j$ and the remaining index $k$, put
\[
 e_{ij}=\frac{b_i+b_j-b_k}{2},\qquad
 u_{ij}=\min(p_i+p_j,2-p_i-p_j),\qquad
 D=\left(u_{12}+u_{13}+u_{23}-2\right)_+.
\]
Then
\begin{equation}
 U=\frac12(\alpha-\beta)^\top p+
       \sum_{i<j}e_{ij}u_{ij}-\min_{i<j}e_{ij}\,D.
 \label{eq:supp-three-formula}
\end{equation}
Zero-mass states with deterministic response zero may be added to apply this expression when $n<3$.

\paragraph{Proof.}
The numbers $b_i$ satisfy the triangle inequalities, hence $e_{ij}\ge0$.
If all $w_i\le1/2$, use $b_i=w_i$ and $w_i+w_j=1-w_k\ge w_k$.
If, for example, $w_1>1/2$, then $b_1=w_2+w_3=b_2+b_3$.
For every binary assignment,
\[
 b_w(S(z))=\sum_{i<j}e_{ij}\mathbf 1\{z_i\ne z_j\}.
\]
It remains to maximize a nonnegative weighted triangle disagreement.

Write $d_{ij}=\mathbb P(Z_i\ne Z_j)$ and $\ell_{ij}=|p_i-p_j|$.
Three Bernoulli variables with means $p_i$ admit specified disagreements exactly when
\[
 \ell_{ij}\le d_{ij}\le u_{ij},\qquad
 d_{ij}\le d_{ik}+d_{jk},\qquad
 d_{12}+d_{13}+d_{23}\le2.
\]
One direct verification uses $q_{ij}=(p_i+p_j-d_{ij})/2$ and the triple intersection probability $t$.
All eight cell probabilities are nonnegative exactly when
\begin{align*}
 t&\geq\max\{0,q_{12}+q_{13}-p_1,q_{12}+q_{23}-p_2,q_{13}+q_{23}-p_3\},\\
 t&\leq\min\{q_{12},q_{13},q_{23},1-\textstyle\sum_i p_i+\sum_{i<j}q_{ij}\}.
\end{align*}
Comparing its lower and upper endpoints gives the displayed pair and triangle inequalities.

The vector $u$ already obeys the three triangle inequalities.
For example, $u_{ik}+u_{jk}$ is the minimum of
$p_i+p_j+2p_k$, $4-p_i-p_j-2p_k$, $2+p_i-p_j$, and $2-p_i+p_j$.
The first two are at least $u_{ij}$; the last two are at least $1\ge u_{ij}$.
If $D=0$, $d=u$ is therefore feasible and optimal.
If $D>0$, any single edge can absorb a reduction of $D$ without crossing its lower bound:
\[
 u_{ik}+u_{jk}+|p_i-p_j|\le2
 \quad\Longrightarrow\quad u_{ij}-\ell_{ij}\ge D.
\]
To verify the first inequality, put $x_i=2p_i-1$ and apply
$|x_i-x_j|\le|x_i+x_k|+|x_j+x_k|$.
Reduce an edge of minimum weight $e_{ij}$ by $D$ and leave the others at $u$.
The resulting vector has sum two and all coordinates in $[0,1]$, so all triangle inequalities hold.
It is feasible and loses exactly $\min e_{ij}\,D$ from the componentwise upper objective.
Every feasible vector must lose at least this amount because its total reduction from $u$ is at least $D$ and all edge weights are at least $\min e_{ij}$.
Combining with~\eqref{eq:supp-three-exact} proves~\eqref{eq:supp-three-formula}.\hfill$\square$

\paragraph{Explicit three-state attaining mechanism.}
Use the preceding optimal $d$, form $q_{ij}$, and choose any $t$ in the nonempty interval above.
Set
\begin{align*}
 Q_{111}&=t,&Q_{110}&=q_{12}-t,\\
 Q_{101}&=q_{13}-t,&Q_{011}&=q_{23}-t,\\
 Q_{100}&=p_1-q_{12}-q_{13}+t,&Q_{010}&=p_2-q_{12}-q_{23}+t,\\
 Q_{001}&=p_3-q_{13}-q_{23}+t,&Q_{000}&=1-\sum_i p_i+\sum_{i<j}q_{ij}-t.
\end{align*}
These masses have the required response means.
Combine this law independently with $C_*$ from~\eqref{eq:supp-min-diag}.
This constructs a shared structural mechanism attaining the upper endpoint.

\section{Equal terminal risks: partition structure, sharp gaps, and computation}
\label{sec:supp-half}
Suppose $p_i=1/2$ for every state. Then
\begin{equation}
 [L,U]=[0,b(w)].
 \label{eq:supp-half}
\end{equation}
\paragraph{Proof.}
The upper bound follows from~\eqref{eq:supp-cut-envelope} because its shift is zero and every cut value is at most $b(w)$.
Choose a maximizing cut $S$, its crossing-maximizing coupling from~\eqref{eq:supp-cut-transport}, and a response law that puts probability $1/2$ on each of $\mathbf1_S$ and $1-\mathbf1_S$.
Its response means are all $1/2$, and its harm is $b(w)$ by~\eqref{eq:supp-disagreement}.
The comonotone all-zero/all-one mixture yields lower endpoint zero.\hfill$\square$

\paragraph{The pairwise endpoint and its worst gap.}
The local pairwise relaxation assigns upper harm $1/2$ to every off-diagonal state pair and zero to a diagonal pair.
Using~\eqref{eq:supp-min-diag}, its upper endpoint is
\[
 U_{\rm pair}=\frac12\left[1-\sum_i(\alpha_i+\beta_i-1)_+\right]
             =\min\left(\frac12,1-\max_iw_i\right).
\]
Its gap satisfies $0\le U_{\rm pair}-U\le1/6$, and the constant is attained for uniform $w$ on three states.
Indeed, if $\max_iw_i\ge1/2$, its complementary subset gives $b(w)=1-\max_iw_i=U_{\rm pair}$.
Otherwise $U_{\rm pair}=1/2$ and a subset with mass in $[1/3,2/3]$ always exists.
Use a single atom in that interval if one exists; otherwise successively add atoms smaller than $1/3$ until reaching $1/3$.
The resulting mass is below $2/3$, proving $b(w)\ge1/3$.
For uniform three-state $w$, $b(w)=1/3$.

\paragraph{Weak NP-completeness.}
Given positive integers $a_1,\ldots,a_n$, take $\alpha=\beta=w$ with $w_i=a_i/\sum_ja_j$.
By~\eqref{eq:supp-half}, $U=1/2$ exactly when the integers have a perfect partition.
Thus deciding whether $U=1/2$ in this restricted two-stage class is NP-hard by PARTITION, and it is in NP because a subset is a certificate.
For this integer-weight input encoding, the problem is weakly NP-complete, with the pseudo-polynomial exact algorithm below.
We make no weak-versus-strong hardness classification for separately encoded rational probabilities.

\paragraph{Exact computation.}
For rational weights $w_i=a_i/W$ with nonnegative integer $a_i$ summing to $W$, the largest achievable subset sum not exceeding $\lfloor W/2\rfloor$ is $WU$.
Standard subset-sum dynamic programming computes it in $O(nW)$ elementary Boolean operations and $O(W)$ memory, with predecessor recovery of an attaining subset.
The cut witness above then yields an attaining structural mechanism.
This uses a classical algorithm; the assertion here is the exact causal reduction.

\paragraph{Certified approximation.}
For additive error $\eta>0$, set $h=\eta/n$, round each weight down to $a_i=\lfloor w_i/h\rfloor$, and solve subset sum up to capacity $\lfloor1/(2h)\rfloor$.
Let $S$ maximize the rounded sum and evaluate its original mass, obtaining $v=\min(w(S),1-w(S))$.
Then
\[
 v\le U\le\min\{U_{\rm pair},v+\eta\}.
\]
To prove this, select an optimal subset $S_*$ with $w(S_*)\le1/2$.
Its rounded sum is feasible, so the selected rounded mass is at least $U-\eta$.
The selected original mass is at least that rounded mass and at most $1/2+\eta$.
Both it and its complement therefore have mass at least $U-\eta$.
The capacity is $O(n/\eta)$, giving $O(n^2/\eta)$ time and $O(n/\eta)$ stored dynamic-programming values under unit-cost arithmetic.
Bit-complexity and witness storage must additionally account for integer word length.

A fully polynomial multiplicative approximation follows as well.
If $\max_iw_i\ge1/2$, compute the exact answer $1-\max_iw_i$.
Otherwise $U\ge1/3$, and additive error $\eta=\epsilon/3$ gives $v\ge(1-\epsilon)U$ in $O(n^2/\epsilon)$ time.
For exact certificates, rounding is performed on integer/rational input, or numerical errors must be included in the interval.

\section{Uniform intermediate states and homogeneous terminal risk}
\label{sec:supp-homogeneous}
If $\alpha=\beta=(1/n,\ldots,1/n)$ and $p_i=p$, then
\begin{equation}
 U=\min\left\{p,1-p,\frac{\lfloor n/2\rfloor}{n}\right\}.
 \label{eq:supp-homogeneous}
\end{equation}
\paragraph{Proof.}
For fixed $Q$, a maximizing first-stage coupling is a scaled permutation matrix.
For a deterministic response assignment, each permutation cycle of length $\ell$ contains at most $\lfloor\ell/2\rfloor$ directed $1\to0$ edges.
The total is at most $\lfloor n/2\rfloor$, and averaging over $Q$ preserves the bound.
The marginal inequalities $H\le p$ and $H\le1-p$ give the remaining upper bounds.

For attainment, take a single directed $n$-cycle as the first-stage permutation.
For integer $m\in\{0,\ldots,n\}$, choose a circular binary pattern with $m$ ones and the maximum number $\min(m,n-m)$ of $1\to0$ edges.
Uniformly mixing its cyclic rotations gives every response coordinate mean $m/n$ and attains~\eqref{eq:supp-homogeneous} at $p=m/n$.
Mix the constructions for $m=\lfloor np\rfloor$ and $\lceil np\rceil$ for arbitrary $p$.
The right side of~\eqref{eq:supp-homogeneous} is linear between consecutive grid points, including the odd-$n$ plateau, so the mixture attains it.
For $n=1$ consistency gives zero directly.\hfill$\square$

\section{Stability in terminal risks}
\label{sec:supp-risk-stability}
For fixed state marginals and any $p,q\in[0,1]^n$,
\begin{equation}
 \left|U(p)-U(q)-\frac12(\alpha-\beta)^\top(p-q)\right|
       \le\sum_iw_i|p_i-q_i|.
 \label{eq:supp-risk-lipschitz}
\end{equation}
\paragraph{Proof.}
Given any response law with means $p$, construct a coupled response vector with means $q$ by coordinate flips.
If $q_i>p_i$, independently flip a zero coordinate to one with conditional probability $(q_i-p_i)/(1-p_i)$.
If $q_i<p_i$, flip a one coordinate to zero with probability $(p_i-q_i)/p_i$.
Endpoint cases with a zero denominator are deterministic.
The resulting coordinate changes with probability exactly $|p_i-q_i|$.
All additional randomness remains independent of the first-stage noise.
For any fixed $C$, the two disagreement indicators can differ only if a response coordinate selected by one of the two state variables changes.
The union bound gives a difference at most
$\sum_i(\alpha_i+\beta_i)|p_i-q_i|=2\sum_iw_i|p_i-q_i|$.
Mapping a maximizer in each direction proves the same Lipschitz bound for the maximal disagreement probability.
Equation~\eqref{eq:supp-disagreement} proves the result.\hfill$\square$

In particular, with $q_i=1/2$, the exact partition endpoint gives the enclosure
\[
 U(p)\in b(w)+\frac12(\alpha-\beta)^\top(p-\tfrac12\mathbf1)
       +\left[-\sum_iw_i|p_i-1/2|,\ \sum_iw_i|p_i-1/2|\right].
\]
Intersect it with $[\max(0,m_A-m_B),\min(m_A,1-m_B)]$.
This encloses the sharp \emph{upper endpoint}; its lower limit is not generally a lower bound on the actual unknown harm.

If $p_i\in[l_i,u_i]$ and $q_i\in[l_i,u_i]$, optimizing~\eqref{eq:supp-risk-lipschitz} coordinatewise also gives
\begin{align*}
 U(p)&\ge U(q)-\sum_i\max\{\alpha_i(q_i-l_i),\beta_i(u_i-q_i)\},\\
 U(p)&\le U(q)+\sum_i\max\{\beta_i(q_i-l_i),\alpha_i(u_i-q_i)\}.
\end{align*}
These are valid risk-box enclosures, not a claim of sharpness over the box.

\section{State-distribution stability and sampling uncertainty}
\label{sec:supp-sampling}
For probability vectors $w,w'$,
\begin{equation}
 |b(w)-b(w')|\le\operatorname{TV}(w,w').
 \label{eq:supp-tv}
\end{equation}
Indeed, every fixed-subset objective is 1-Lipschitz in its subset mass, and
$\operatorname{TV}(w,w')=\max_S|w(S)-w'(S)|$; taking maxima preserves the bound.

Suppose independent randomized samples give $N_A$ state observations from $\alpha$ and $N_B$ from $\beta$, and let $\widehat w=(\widehat\alpha+\widehat\beta)/2$.
Conditional on the stipulated known terminal risks $p_i=1/2$, and for failure probability $\zeta\in(0,1)$, set
\[
 r_\zeta=\sqrt{\frac{1/N_A+1/N_B}{8}\log\frac{2^n}{\zeta}}.
\]
For a fixed subset $S$, weighted Hoeffding concentration yields
\[
 \mathbb P\{\widehat w(S)-w(S)\ge r\}
       \le\exp\left\{-\frac{8r^2}{1/N_A+1/N_B}\right\}.
\]
Since total variation is the maximum \emph{one-sided} subset difference, a union bound over $2^n$ subsets gives
$\mathbb P\{\operatorname{TV}(\widehat w,w)\le r_\zeta\}\ge1-\zeta$.
If dynamic programming returns certified endpoints
$L_{\rm DP}\le b(\widehat w)\le U_{\rm DP}$, then
\[
 \left[\max(0,L_{\rm DP}-r_\zeta),\ \min(1/2,U_{\rm DP}+r_\zeta)\right]
\]
is a confidence interval for the sharp upper endpoint $b(w)$.
An uncertainty interval for the actual unknown harm instead uses zero as its lower limit.
When $N_A=N_B=N$, the empirical pooled weights equal
$(\mathrm{count}_{A,i}+\mathrm{count}_{B,i})/(2N)$, enabling exact integer subset-sum computation with capacity $N$.
Combining uncertain state laws with uncertain terminal risks requires a separate simultaneous argument and is not claimed by this conditional guarantee.

\section{A four-state cut-envelope counterexample}
\label{sec:supp-four}
For
\[
 \alpha=(1,1,4,4)/10,\quad \beta=(4,1,4,1)/10,\quad
 p=(3,2,2,1)/4,
\]
exhaustive vertex optimization gives $U=3/10$, whereas the cut envelope and rectangular bound both equal $7/20$.
A mechanism attaining $3/10$ has
\[
 C=\begin{pmatrix}
 0&1/10&0&0\\
 0&0&0&1/10\\
 4/10&0&0&0\\
 0&0&4/10&0
 \end{pmatrix}
\]
and puts response probability $1/4$ on each of $0110,1001,1010,1100$.
The exact rational upper certificate enumerates the transportation and response polytope vertices using integer fraction-free elimination; some maximizer of a bilinear objective on two compact polytopes is attained at a vertex pair.
There are 68 distinct transport vertices and 58 distinct response vertices for these rational inputs.
The accompanying verification inspects all $\binom{16}{7}=11{,}440$ transportation bases and all $\binom{16}{5}=4{,}368$ response bases, compares all $68\times58=3{,}944$ distinct feasible vertex pairs, and checks primal feasibility in exact rational arithmetic. No floating-point feasibility or rank decisions enter this certificate.
This example establishes the limit of three-state envelope exactness; it does not establish a new general marginal-polytope result.

\section{A dominant pooled state makes the pairwise bound sharp}
\label{sec:supp-dominant}
Under the same two-stage binary-outcome model, suppose there is a state $k$ with
$w_k=(\alpha_k+\beta_k)/2\ge1/2$.
Then, for every heterogeneous risk vector $p\in[0,1]^n$,
\begin{equation}
 U=U_{\rm pair}
  =\sum_{i\ne k}\left[\alpha_i\min(p_i,1-p_k)
                    +\beta_i\min(p_k,1-p_i)\right].
 \label{eq:supp-dominant}
\end{equation}
This result has no restriction on $n$ and does not require balanced terminal risks.
It is a sufficient condition for sharpness, not a necessary one.

\paragraph{Proof: the pairwise transport optimum is a star.}
Write $h_{ij}=\min(p_i,1-p_j)$ for $i\ne j$ and $h_{ii}=0$.
For distinct $i,j,k$,
\begin{equation}
 \min(p_i,1-p_j)\le
 \min(p_i,1-p_k)+\min(p_k,1-p_j).
 \label{eq:supp-harm-triangle}
\end{equation}
To verify this inequality, expand its right side as
\[
 \min\{p_i+p_k,\ p_i+1-p_j,\ 1,\ 2-p_j-p_k\}.
\]
Each of these four quantities is at least $\min(p_i,1-p_j)$.
For $i=j\ne k$, the required cost inequality is simply
$h_{ii}=0\le h_{ik}+h_{ki}$.

Let $C\in\Pi(\alpha,\beta)$ be arbitrary, and put
$d=\alpha_k+\beta_k-1\ge0$.
Its hub diagonal satisfies
\[
 C_{kk}=d+\sum_{i,j\ne k}C_{ij}.
\]
For every entry $C_{ij}$ outside the hub row and column, including a non-hub diagonal entry, reroute an amount $x=C_{ij}$ as follows:
remove $x$ from $(i,j)$ and $(k,k)$, and add $x$ to $(i,k)$ and $(k,j)$.
Row and column marginals are preserved. The displayed identity guarantees enough hub-diagonal mass for all reroutings, and~\eqref{eq:supp-harm-triangle} guarantees that the pairwise objective cannot decrease.
The resulting coupling is
\[
 C^*_{ik}=\alpha_i\ (i\ne k),\qquad
 C^*_{kj}=\beta_j\ (j\ne k),\qquad
 C^*_{kk}=d,
\]
with all remaining entries zero. It is consequently a pairwise transport maximizer and its objective is the right side of~\eqref{eq:supp-dominant}.

\paragraph{Proof: one shared response vector attains the star.}
Independently of the first-stage draw from $C^*$, draw $V\sim\operatorname{Uniform}(0,1)$ and define
\[
 Z_k=\mathbf1\{V\le p_k\},\qquad
 Z_i=\mathbf1\{V>1-p_i\}\quad(i\ne k).
\]
Every response coordinate has its prescribed mean $p_i$.
For every non-hub state $i$ simultaneously,
\[
 \mathbb P(Z_i=1,Z_k=0)=\min(p_i,1-p_k),\qquad
 \mathbb P(Z_k=1,Z_i=0)=\min(p_k,1-p_i).
\]
Thus one response law attains all the pairwise maxima used by $C^*$; it also respects identical-state consistency on $(k,k)$.
It realizes the pairwise upper endpoint with a legal shared SCM, so $U\ge U_{\rm pair}$.
The converse inequality holds because the pairwise problem is an outer relaxation.
This proves~\eqref{eq:supp-dominant}. The $n=1$ case gives zero by the empty sum.\hfill$\square$

\paragraph{Assumptions and interpretation.}
The only assumptions are those already stated for the main model: binary terminal outcomes, one common response mechanism, and independence of the first-stage and response-vector noises.
No ordering, monotonicity, positivity of the risk values, or equality of terminal risks is added.
The theorem also covers $w_k=1/2$ and zero-mass states.
It provides a conditional explanation for native simulator cases with a dominant pooled state; without that mass condition, a zero observed gap can still occur, and this result makes no converse claim.

\section{Extension models and information sets}
\label{sec:ext-models}
Let $\alpha,\beta$ be the two first-stage state laws and $\Pi(\alpha,\beta)$ their coupling polytope.
For a shared binary response vector with means $p$, write $L(\alpha,\beta,p)$ and $U(\alpha,\beta,p)$ for the independent-stage sharp harm endpoints established in the main paper.
Their centered versions are
\[
 E_L(\alpha,\beta,p)=L(\alpha,\beta,p)-\tfrac12(\alpha-\beta)^\top p,
 \qquad
 E_U(\alpha,\beta,p)=U(\alpha,\beta,p)-\tfrac12(\alpha-\beta)^\top p.
\]
They are one half of the smallest and largest cross-world disagreement probabilities.
Throughout, $\operatorname{TV}(P,Q)=\sup_A|P(A)-Q(A)|=\tfrac12\|P-Q\|_1$.

For direct effects, use a joint response law $Q$ on $(Z^A,Z^B)\in\{0,1\}^{2n}$ with
\begin{equation}
 \mathbb E_QZ_i^A=a_i,\qquad \mathbb E_QZ_i^B=b_i,
 \qquad \mathbb P_Q(Z_i^A\ne Z_i^B)\le\delta_i.
 \label{eq:ext-direct-constraints}
\end{equation}
The vectors $a,b$ are state-intervention response risks, and $\delta_i\in[0,1]$ are declared sensitivity budgets.
A dependence extension permits a joint law $J$ of $(S^A,S^B,Z^A,Z^B)$ with marginals $C$ and $Q$, subject to
\begin{equation}
 C\in\Pi(\alpha,\beta),\qquad
 \operatorname{TV}(J,C\otimes Q)\le\rho.
 \label{eq:ext-tv}
\end{equation}
The observed regime risks $m_A=\mathbb P_J(Z^A_{S^A}=1)$ and $m_B=\mathbb P_J(Z^B_{S^B}=1)$, when supplied, are separate primitives.
Under dependence they need not equal $\alpha^\top a$ and $\beta^\top b$.
Every theorem below is conditional on a nonempty stated model class.
The budgets are sensitivity parameters, not quantities identified by the transition marginals alone.

\section{Stage dependence and mean calibration}
\label{sec:ext-dependence}
First retain one shared response vector $Z$ with means $p$.
For any $J$ satisfying~\eqref{eq:ext-tv}, its product-marginal counterpart $P_0=C\otimes Q$ is an admissible independent-stage model.
Since harm is an event,
\[
 H(J)\in[L(\alpha,\beta,p)-\rho,\ U(\alpha,\beta,p)+\rho].
\]
This elementary bound does not imply that the regime means retain their product-model values.

\paragraph{Proposition: calibrated dependence bound.}
If $m_A,m_B$ are also fixed, define
\[
 \tau_p=\tfrac12\{m_A-m_B-(\alpha-\beta)^\top p\}.
\]
Then
\begin{equation}
 \begin{aligned}
 H(J)\in{}&[L(\alpha,\beta,p)+\tau_p-\rho/2,
              U(\alpha,\beta,p)+\tau_p+\rho/2]\\
            &\cap[\max(0,m_A-m_B),\min(m_A,1-m_B)].
 \end{aligned}
 \label{eq:ext-calibrated-tv}
\end{equation}
\paragraph{Proof.}
For any two binary variables $A,B$,
$\mathbb P(A=1,B=0)=\{\mathbb EA-\mathbb EB+\mathbb P(A\ne B)\}/2$.
The disagreement event under $J$ and $P_0$ differs in probability by at most $\rho$.
Their mean differences differ by $2\tau_p$.
Apply the identity and the independent-stage endpoint bounds; the second interval is the ordinary endpoint Fr\'echet restriction.\hfill$\square$

Even the calibrated interval is not generally sharp: with a single shared state, consistency forces harm to zero for every admissible law.
If an observed dataset fixes the full $(S,Y)$ table under each regime, those tables can be imposed in addition to the means; all bounds above remain valid.

\paragraph{Unrestricted dependence is a finite sharp LP.}
With no restriction~\eqref{eq:ext-tv}, variables $J_{ijz}\ge0$ for a shared response vector obey the linear constraints
\[
 \sum_{j,z}J_{ijz}=\alpha_i,\quad
 \sum_{i,z}J_{ijz}=\beta_j,\quad
 \sum_{i,j,z}J_{ijz}z_s=p_s.
\]
The harm objective is $\sum_{i,j,z}J_{ijz}z_i(1-z_j)$.
Known regime risks or full state--outcome tables add linear equalities, for example
$\sum_{j,z}J_{ijz}z_i=\alpha_i p_i$ when that particular table is part of the information set.
Every feasible joint law is realized by one common latent variable indexing $(i,j,z)$ and the corresponding structural lookup functions.
Thus minimizing and maximizing this LP gives sharp bounds for precisely its stated information set.
There are $n^2 2^n$ variables.
In contrast, a nonzero TV restriction with unknown $C,Q$ contains the product $C_{ij}Q_z$ and is not generally a single LP. Fixing $C,Q$ makes its absolute-deviation formulation linear, but solves a smaller problem.

\section{Bounded direct effects with independent stages}
\label{sec:ext-direct}
Consider $J=C\otimes Q$ with~\eqref{eq:ext-direct-constraints}.
The response constraints are feasible exactly when $|a_i-b_i|\le\delta_i$ for every state.
Necessity follows from the difference between Bernoulli means; sufficiency follows by taking
$Z_i^A=\mathbf1\{V\le a_i\}$ and $Z_i^B=\mathbf1\{V\le b_i\}$ for one independent uniform $V$.

\paragraph{Theorem: the sharp lower endpoint survives direct-effect budgets.}
For every feasible budget vector,
\begin{align}
 L_{\rm dir}
 &=\min_{C\in\Pi(\alpha,\beta)}\sum_{i,j}C_{ij}(a_i-b_j)_+\\
 &=\tfrac12\left\{W_1\left(\sum_i\alpha_i\delta_{a_i},\sum_j\beta_j\delta_{b_j}\right)
                       +\alpha^\top a-\beta^\top b\right\}.
 \label{eq:ext-direct-lower}
\end{align}
\paragraph{Proof.}
Every response law has cross-state harm at least $(a_i-b_j)_+$.
Here $\delta_x$ denotes a Dirac mass at $x$, distinct from the sensitivity budget $\delta_i$.
The common forward-threshold construction just given simultaneously attains all these lower bounds, and its within-state disagreement is exactly $|a_i-b_i|$, respecting every feasible budget.
Minimize over $C$ and use $(x-y)_+=(|x-y|+x-y)/2$.\hfill$\square$

In particular, when $a=b=p$, the original sharp lower endpoint is unchanged.
This conclusion does not hold if the second risk vector is left unrestricted rather than fixed.

\paragraph{Proposition: direct-effect upper certificates.}
For a direct-effect model, replace the $B$-response by $Z^A$ to obtain an admissible shared-response anchor with risks $a$.
Its disagreement event can differ only when $Z^A_{S^B}\ne Z^B_{S^B}$, whose probability is at most $\beta^\top\delta$ by stage independence.
Consequently
\begin{equation}
 U_{\rm dir}\le
 U(\alpha,\beta,a)+\tfrac12\beta^\top(a-b)+\tfrac12\beta^\top\delta.
 \label{eq:ext-direct-upper-a}
\end{equation}
The analogous $b$-anchor gives
\begin{equation}
 U_{\rm dir}\le
 U(\alpha,\beta,b)+\tfrac12\alpha^\top(a-b)+\tfrac12\alpha^\top\delta.
 \label{eq:ext-direct-upper-b}
\end{equation}
Both may be intersected with the endpoint Fr\'echet upper bound.
When $a=b=p$, all original shared-response models remain feasible, so
$U(\alpha,\beta,p)\le U_{\rm dir}\le U(\alpha,\beta,p)+\min(\alpha^\top\delta,\beta^\top\delta)/2$.
These are valid general certificates, not unrestricted sharp formulas.

The direct-effect response polytope has $4^n$ cells with linear mean and mismatch constraints.
For fixed $C$, its endpoint problem is an LP.
For unknown $C$, the objective remains bilinear; enumerating every transportation vertex and solving one response LP per vertex yields a globally sharp finite procedure.
Indeed, at any global extremizer, hold its response law fixed and replace $C$ by a transportation vertex attaining an at least equally extreme linear objective.
Reoptimizing the response LP at that vertex recovers the global endpoint.
The product of these two nonempty polytopes is path connected, so all values between its extrema are attained.

\section{Joint relaxation of independence and full mediation}
\label{sec:ext-joint}
Let~\eqref{eq:ext-direct-constraints} and~\eqref{eq:ext-tv} hold, with known regime risks $m_A,m_B$.
Set
\[
 \begin{aligned}
 \tau_a&=\tfrac12\{m_A-m_B-(\alpha-\beta)^\top a\},\\
 \tau_b&=\tfrac12\{m_A-m_B-(\alpha-\beta)^\top b\},\\
 \tau_{\rm dir}&=\tfrac12\{m_A-m_B-\alpha^\top a+\beta^\top b\}.
 \end{aligned}
\]
Then a valid joint outer interval has lower endpoint
\begin{equation}
 \max\{0,m_A-m_B,L_{\rm dir}+\tau_{\rm dir}-\rho/2\}
 \label{eq:ext-joint-lower}
\end{equation}
and upper endpoint
\begin{equation}
 \min\left\{m_A,1-m_B,
 U(\alpha,\beta,a)+\tau_a+\tfrac12(\rho+\beta^\top\delta),
 U(\alpha,\beta,b)+\tau_b+\tfrac12(\rho+\alpha^\top\delta)\right\}.
 \label{eq:ext-joint-upper}
\end{equation}
If an exact or certified $U_{\rm dir}$ is available, also intersect with
$U_{\rm dir}+\tau_{\rm dir}+\rho/2$.

\paragraph{Proof.}
Compare the disagreement event under $J$ with its product-marginal direct-effect model $C\otimes Q$; this costs at most $\rho$ in disagreement probability.
The product model has the exact lower bound~\eqref{eq:ext-direct-lower}.
For an upper certificate compare its disagreement with the $a$-anchor or $b$-anchor, costing respectively at most $\beta^\top\delta$ or $\alpha^\top\delta$.
Apply the binary harm/disagreement identity, retaining the fixed observed mean difference.
This gives~\eqref{eq:ext-joint-lower}--\eqref{eq:ext-joint-upper}.\hfill$\square$

A symmetric anchor-based lower bound is also valid, but generally weaker than~\eqref{eq:ext-joint-lower}.
The exact independent-stage lower endpoint does not remain unchanged after arbitrary stage dependence is added.
These joint certificates are outer bounds; no general sharpness or convexity claim is made for the unknown-product TV model.

\section{A sharp two-parameter sensitivity surface}
\label{sec:ext-surface}
Consider three uniform first-stage states, state-intervention risks
$a=b=(1/2,1/2,1/2)$, known regime risks $m_A=m_B=1/2$, and homogeneous direct budgets $\delta_i=\delta$.
The dependence budget is $\rho$.
Then the full attainable harm interval is
\begin{equation}
 \left[0,\ \min\left\{\tfrac12,\tfrac13+\tfrac12(\rho+\delta)\right\}\right].
 \label{eq:ext-surface}
\end{equation}
The upper certificate follows immediately from~\eqref{eq:ext-joint-upper} and the original three-state bound $1/3$.
We next give an attaining construction that preserves even the two full state--outcome tables.

\paragraph{The independent direct-effect path.}
Fix the uniform cyclic coupling $C_{i,\sigma(i)}=1/3$, where $\sigma=(1\ 2\ 3)$.
Let $\mathcal F$ be the six nonconstant binary vectors in $\{0,1\}^3$.
Under $Q_0$, draw $A$ uniformly from $\mathcal F$ and set $B=A$.
Under $Q_1$, draw the same $A$ and define $B_{\sigma(i)}=1-A_i$.
Both response vectors have coordinate means $1/2$.
Under $C\otimes Q_0$, harm equals $1/3$ and every statewise direct discrepancy is zero.
Under $C\otimes Q_1$, the two queried responses always differ, so harm equals $1/2$.
Moreover, $A_i\ne B_i$ exactly when $A_i=A_{\sigma^{-1}(i)}$.
Every vector in $\mathcal F$ has exactly one equal edge around the three-cycle, so rotational symmetry gives statewise discrepancy $1/3$.
The paired response supports of $Q_0$ and $Q_1$ are disjoint: equality would require alternating bits around an odd cycle.

Set $d=\min(\delta,1/3)$, $\lambda=3d$, and $Q_d=(1-\lambda)Q_0+\lambda Q_1$.
Its direct discrepancy at every state is $d$, and the independent product has harm $1/3+d/2$.

\paragraph{Adding dependence without changing observed marginals.}
In the $Q_0$ part of $C\otimes Q_d$, each response column has one $10$ edge, one $01$ edge, and one equal edge; each such joint cell has mass $(1-\lambda)/18$.
There is total equal-cell mass $(1-\lambda)/3=1/3-d$.
Put $r=\min(\rho,1/3-d)$.
If this available mass is positive, remove the same fraction of each equal cell, moving total mass $r$; within each response column split the removed mass equally between its $10$ and $01$ cells.
If there is no available mass, take $r=0$.
Column masses, and hence $Q_d$, are unchanged.
Rotational and complement symmetry preserve each cyclic row mass and each row's $A$- and $B$-response mean $1/2$.
Thus $C$, $Q_d$, both full $(S,Y)$ tables, and the direct budgets are preserved.
The total variation from the product-marginal law is exactly $r$, and the harm increases by $r/2$.
The final value is $1/3+(d+r)/2$, equal to the upper endpoint in~\eqref{eq:ext-surface}.

To obtain the interval below $1/3$, retain independent stages and mix $Q_0$ with the common all-zero/all-one response law, which has harm zero and the same marginal risks.
For the portion above $1/3$, vary the used direct and dependence budgets continuously within the allowed budgets in the construction above.
Hence every value in~\eqref{eq:ext-surface} is attained; this conclusion uses the explicit construction, not an unproved convexity claim about the TV-constrained model.\hfill$\square$

Setting $\delta=0$ or $\rho=0$ gives the separate sharp dependence and direct-effect frontiers.
The surface is a specialized calibration result, not a formula for arbitrary primitive risks or state distributions.

\section{Multistage common care: a sharp restricted extension and a folding failure}
\label{sec:ext-multistage}
For a common future policy with future noises independent of the preceding block, one can define $Z_i$ to be the final outcome when the common continuation starts in state $i$.
This gives a valid response-vector representation, but known internal transition kernels can restrict its feasible law below the full set of joint Bernoulli vectors with the same terminal marginals.
Known earlier transitions can similarly restrict the first-stage coupling.
Consequently applying the unrestricted folded model yields an outer interval in general.
It is sharp for a queried endpoint only if at least one folded extremizer is realizable by all retained internal kernels.
For example, folding is exact for a free first-stage coupling followed by $S\to T\to Y=T$, where $T$ is binary and arbitrary shared response laws for the specified $S\to T$ marginals are allowed.

\paragraph{Strictly positive three-stage counterexample.}
Take
\[
 \begin{aligned}
 \alpha=\beta&=(1/5,3/10,1/2),\\
 \mathbb P(T=1\mid S)&=(19/20,1/20,19/20),\\
 \mathbb P(Y=1\mid T=0)&=\mathbb P(Y=1\mid T=1)=1/2.
 \end{aligned}
\]
All transition entries are strictly positive and noises at the three stages are independent.
The folded terminal risks are all $1/2$, so its upper endpoint is $1/2$.
Write the last-stage response pair as $(R_0,R_1)$.
The final disagreement event is exactly
$\{T^A\ne T^B\}\cap\{R_0\ne R_1\}$.
Independence factors its probability, and the second factor can attain one by choosing complementary half-risk responses.
Since the two $T$ marginal risks agree, the true final upper endpoint equals the original three-state upper formula with $T$ as its binary endpoint.
That formula has edge weights $(e_{12},e_{13},e_{23})=(0,1/5,3/10)$ and pairwise disagreement bounds $(1,1/10,1)$, giving
\[
 U_{\rm true}=\frac15\frac1{10}+\frac3{10}=\frac8{25}<\frac12=U_{\rm folded}.
\]
A common last-stage response $R_0=R_1$ attains lower harm zero.
This is a fixed-kernel sharp interval $[0,8/25]$, not a deterministic-erasure example.

\paragraph{Theorem: a three-state entrance followed by a binary Markov suffix.}
Suppose the first two stages give a binary state $T$ after a three-state mediator, so the main theorem provides exact bounds $[L_T,U_T]$.
Let $\Delta_T=\mathbb P(T^A=1)-\mathbb P(T^B=1)$ and $D_T^-=2L_T-\Delta_T$, $D_T^+=2U_T-\Delta_T$.
Consider any finite subsequent sequence of common binary Markov transitions with row risks $(r_{t0},r_{t1})$ and independent time-specific noises.
The common policy may depend on time and the current binary state; it must not choose different actions for equal current states based on unrecorded different histories.
Put
\[
 a_t=|r_{t0}-r_{t1}|,\qquad
 b_t=\min(r_{t0}+r_{t1},2-r_{t0}-r_{t1}),\qquad
 \Delta_f=\Delta_T\prod_t(r_{t1}-r_{t0}).
\]
Then the final sharp endpoints are
\begin{equation}
 L_f=\tfrac12\left\{\Delta_f+D_T^-\prod_ta_t\right\},\qquad
 U_f=\tfrac12\left\{\Delta_f+D_T^+\prod_tb_t\right\}.
 \label{eq:ext-binary-suffix}
\end{equation}
\paragraph{Proof.}
At each binary step, equal counterfactual states remain equal under the common structural response.
Different states remain different exactly when that step's two response entries differ.
Their disagreement probability $d_t$ ranges over $[a_t,b_t]$ by the Bernoulli coupling bounds.
Independence from previous noises yields $D_f=D_T\prod_td_t$.
The difference in ordinary marginal risks evolves as
$\Delta_{t+1}=(r_{t1}-r_{t0})\Delta_t$.
All disagreement factors are nonnegative, and their endpoints can be selected independently at different times.
Combine prefix and suffix extremizers, then apply the harm/disagreement identity to prove both attainment and the endpoint formulas.\hfill$\square$

The binary-chain product mechanism has close prior precedents; the claimed scope is the connection to the solved three-state entrance, not a general multivalued multistage solution.

\section{Simultaneous finite-sample outer intervals beyond balanced risks}
\label{sec:ext-finite-sample}
The following stability result holds for either endpoint $V\in\{L,U\}$ of the original independent shared-response model, with $E_V=V-(\alpha-\beta)^\top p/2$.
For any two primitive triples $\theta=(\alpha,\beta,p)$ and $\widehat\theta=(\widehat\alpha,\widehat\beta,\widehat p)$,
\begin{equation}
 |E_V(\theta)-E_V(\widehat\theta)|
 \le\tfrac12\{\operatorname{TV}(\alpha,\widehat\alpha)
               +\operatorname{TV}(\beta,\widehat\beta)\}
       +\sum_i\widehat w_i|p_i-\widehat p_i|,
 \qquad \widehat w=(\widehat\alpha+\widehat\beta)/2.
 \label{eq:ext-centered-stability}
\end{equation}

\paragraph{Proof.}
For fixed response risks, take any coupling $C$ of the original state marginals.
Use maximal-coupling transition kernels to map its first coordinate to marginal $\widehat\alpha$ and second coordinate to $\widehat\beta$.
Use additional first-stage randomness independent of the response vector.
The probability that either selected state changes is at most the sum of the two TV distances, so the disagreement event changes by at most that amount.
Mapping feasible mechanisms in both directions bounds both the minimum and maximum disagreement objectives.
After changing the state marginals, couple response vectors by coordinate flips with unconditional probabilities $|p_i-\widehat p_i|$.
For example, to increase a mean from $p_i$ to $\widehat p_i$, flip a zero with conditional probability $(\widehat p_i-p_i)/(1-p_i)$; use the analogous flip of ones to decrease a mean.
The endpoint cases have the evident deterministic or zero-flip conventions.
These additional random variables belong to the second-stage noise block and remain independent of the state pair.
The disagreement objective changes by at most
$\sum_i(\widehat\alpha_i+\widehat\beta_i)|p_i-\widehat p_i|$.
Divide by two and combine the two comparisons.\hfill$\square$

The constants in this centered bound cannot generally be improved.
For the state term, use two deterministic risk values $(0,1)$ and change one arm's state from the first state to the second: disagreement changes by one when the sum of TV distances is one.
For the risk term, place the two arms on distinct states and change both risk values from zero to $\epsilon\le1/2$; maximum disagreement rises from zero to $2\epsilon$, attaining the weighted-risk term.

\paragraph{Confidence ingredients.}
Suppose $N_A,N_B$ independent-within-arm randomized state observations estimate $\alpha,\beta$.
For prescribed failure allocations $\gamma_A,\gamma_B$, valid TV radii are
\[
 r_A=\min\left\{1,\sqrt{\frac{\log(2^n/\gamma_A)}{2N_A}}\right\},\qquad
 r_B=\min\left\{1,\sqrt{\frac{\log(2^n/\gamma_B)}{2N_B}}\right\}.
\]
They follow from one-sided Hoeffding bounds for each subset and a union bound.
Suppose also that $M_i>0$ independent Bernoulli state-intervention observations estimate each $p_i$.
A simultaneous risk radius is
\[
 \epsilon_i=\min\left\{1,\sqrt{\frac{\log(2n/\gamma_p)}{2M_i}}\right\}.
\]
Clipping the coordinate intervals to $[0,1]$ can improve them.
With $M_i=0$, choose any $\widehat p_i\in[0,1]$ and use
$\epsilon_i=\max(\widehat p_i,1-\widehat p_i)$, giving the vacuous interval $[0,1]$ rather than inferring an unobserved primitive.
A union bound gives simultaneous coverage at least $1-\gamma_A-\gamma_B-\gamma_p$; independence between the different confidence events is not required.
The sampling assumptions concern randomized or otherwise justified interventional primitives, not arbitrary confounded observational conditionals.

\paragraph{A computable confidence outer interval without separately measured means.}
Let $\widehat L\le L(\widehat\theta)$ and $\widehat U\ge U(\widehat\theta)$ be exact endpoints or certified endpoint bounds.
On the simultaneous event just described, the entire true identified interval is contained in
\begin{equation}
 [\widehat L-R,\widehat U+R]\cap[0,1],\qquad
 R=r_A+r_B+\sum_i\max(\widehat\alpha_i,\widehat\beta_i)\epsilon_i.
 \label{eq:ext-finite-simple}
\end{equation}
\paragraph{Proof.}
The difference between the true and estimated marginal-risk contrasts is at most
$r_A+r_B+\sum_i|\widehat\alpha_i-\widehat\beta_i|\epsilon_i$.
Combine half of this bound with~\eqref{eq:ext-centered-stability} and use
$\widehat w_i+|\widehat\alpha_i-\widehat\beta_i|/2
=\max(\widehat\alpha_i,\widehat\beta_i)$.
Certified rather than exact estimated endpoints preserve the outer guarantee.\hfill$\square$

\paragraph{A mean-calibrated confidence outer interval.}
If separately observed regime outcomes supply simultaneous confidence intervals
$m_A\in[\ell_A,u_A]$ and $m_B\in[\ell_B,u_B]$, allocate their failure probabilities in the same overall union bound.
For Bernoulli samples one may use the usual two-sided Hoeffding intervals.
Set $\Delta_-=\ell_A-u_B$, $\Delta_+=u_A-\ell_B$,
$\widehat\Delta=(\widehat\alpha-\widehat\beta)^\top\widehat p$, and
$r_E=(r_A+r_B)/2+\sum_i\widehat w_i\epsilon_i$.
Then the true identified interval is contained in
\begin{equation}
 \left[\widehat L+\frac{\Delta_--\widehat\Delta}{2}-r_E,
       \widehat U+\frac{\Delta_+-\widehat\Delta}{2}+r_E\right]
 \cap[\max(0,\Delta_-),\min(u_A,1-\ell_B)].
 \label{eq:ext-finite-calibrated}
\end{equation}
This is an outer confidence interval for the whole partially identified range, not an estimator of the unknown actual harm or a sharp robust optimization over the nuisance confidence set.
It can improve precision when the regime means are well estimated, but it does not uniformly dominate~\eqref{eq:ext-finite-simple}; in the original model one may intersect the two valid intervals.
The bounds remain valid for arbitrary state counts when $\widehat U$ is a valid outer certificate; exact estimation of the nuisance-specific upper endpoint is not assumed.

\paragraph{Combining confidence and joint sensitivity conservatively.}
The same argument can be applied to the $a$-anchor of~\eqref{eq:ext-joint-upper}.
Let $\widehat L_a,\widehat U_a$ be certified original-model endpoints at
$(\widehat\alpha,\widehat\beta,\widehat a)$, let
$\widehat\Delta_a=(\widehat\alpha-\widehat\beta)^\top\widehat a$, and define
\[
 r_{E,a}=(r_A+r_B)/2+\sum_i\widehat w_i\epsilon_{a,i},\qquad
 B_\delta=\min\{\max_i\delta_i,
       \widehat\beta^\top\delta+r_B(\max_i\delta_i-\min_i\delta_i)\}.
\]
The latter bounds $\beta^\top\delta$ uniformly on the state-law TV ball.
A valid joint confidence-and-sensitivity outer interval is
\[
 \left[\widehat L_a+\frac{\Delta_--\widehat\Delta_a}{2}
                    -r_{E,a}-\frac{\rho+B_\delta}{2},\quad
       \widehat U_a+\frac{\Delta_+-\widehat\Delta_a}{2}
                    +r_{E,a}+\frac{\rho+B_\delta}{2}\right],
\]
intersected with the endpoint restriction in~\eqref{eq:ext-finite-calibrated}.
This symmetric anchor certificate is generally weaker on the lower side than exploiting the exact direct-effect lower transport result.
It requires simultaneous coverage for the state laws, the anchor state-intervention risks, and the regime means; budget values themselves remain declared assumptions.
If an interval intersection above is empty for otherwise valid samples, return $[0,1]$.
Such conflicts can occur off the simultaneous coverage event; this fallback makes the procedure defined for every valid sample and preserves its coverage guarantee.
For homogeneous $\delta_i=\delta$, $B_\delta=\delta$ exactly, so state-law uncertainty does not inflate that budget term.

\section{Simultaneous confidence projection for the simulator audit}
\label{sec:ext-simulator-confidence}
This section concerns a specified simulator SCM that exposes complete response vectors for independently simulated patients.
Such response observations are simulator instrumentation, not ordinary patient observations or identification of counterfactual responses from clinical records.
Write $s$ for the total number of states, including any known absorbing state.
For one fixed simulator setting, suppose $N$ independent patients provide joint counts for $(S^A,S^B,Z)$.
Binomial confidence intervals obtained by exact tail inversion are applied to each of the $s$ coordinates of $\alpha$, $\beta$, and $p$, the two regime means, and a rare-response probability $\varepsilon$.
There are $3s+3$ intervals, with a failure allocation of $\eta/(3s+3)$ for each.
The union bound gives simultaneous coverage of all these primitives with probability at least $1-\eta$; overlap between the underlying observations does not invalidate this statement.
Known deterministic coordinates can be set to their exact values without spending additional error probability.

\paragraph{Rare-response dependence bound, including the adverse-event absorber.}
Let $z_0$ be any fixed complete response vector and let
$\varepsilon=\mathbb P(Z\ne z_0)$.
For $J$ with marginals $C,Q$,
\[
 \operatorname{TV}(J,C\otimes Q)
 \le\operatorname{TV}(J,C\otimes\delta_{z_0})
    +\operatorname{TV}(C\otimes\delta_{z_0},C\otimes Q)
 =2\varepsilon.
\]
To check the first equality, the mass outside $z_0$ is $\varepsilon$, and the mass deficit at $z_0$, summed over state pairs, is also $\varepsilon$; the half-$L^1$ convention therefore gives $\varepsilon$.
The second distance equals $\operatorname{TV}(\delta_{z_0},Q)=\varepsilon$.
In the adverse-event audit, the reference is
$z_0=(0,\ldots,0,1)$: all active-state responses have no adverse event, while the already-adverse-event absorber deterministically retains outcome one.
The rare-response event is that at least one active-state response differs from zero; the absorber is not discarded.
An upper confidence bound $\varepsilon_+$ thus gives
$\rho_+=\min(1,2\varepsilon_+)$.

\paragraph{Linear projection of all primitive confidence boxes.}
Let $\alpha_i\in[\alpha_i^-,\alpha_i^+]$, and define analogous boxes for $\beta,p,m_A,m_B$.
Let $\mathcal T$ contain nonnegative matrices of total mass one whose row and column sums lie in the respective $\alpha$ and $\beta$ boxes.
Set
\[
 c_{ij}^-=(p_i^--p_j^+)_+,\qquad
 c_{ij}^+=\min(p_i^+,1-p_j^-),\qquad i\ne j,
 \qquad c_{ii}^-=c_{ii}^+=0,
\]
and compute
\[
 \ell=\min_{C\in\mathcal T}\langle C,c^-\rangle,
 \qquad u=\max_{C\in\mathcal T}\langle C,c^+\rangle.
\]
On the simultaneous coverage event, the true fixed-marginal transportation polytope is contained in $\mathcal T$, and each true pairwise harm lies between its displayed cost bounds.
Consequently $\ell\le L(\alpha,\beta,p)$ and $U(\alpha,\beta,p)\le u$.
The upper comparison is through the pairwise relaxation and does not require that independent pairwise extremizers be globally compatible.

Let $A^-,A^+$ be the minimum and maximum of $\alpha^\top p$ over the primitive boxes and the probability simplex for $\alpha$, and define $B^-,B^+$ analogously.
These extrema are linear programs: nonnegative probability weights permit setting $p=p^-$ for the minimum and $p=p^+$ for the maximum.
Separate optimization of the two products can be conservative but bounds their true values simultaneously.
Thus
\[
 \tau^- =\tfrac12(m_A^--m_B^+-A^++B^-),\qquad
 \tau^+ =\tfrac12(m_A^+-m_B^--A^-+B^+)
\]
bound the true calibration shift.
By~\eqref{eq:ext-calibrated-tv}, a valid population harm confidence outer interval is
\[
 \left[\max\{0,m_A^--m_B^+,\ell+\tau^- -\rho_+/2\},\quad
       \min\{m_A^+,1-m_B^-,u+\tau^+ +\rho_+/2\}\right].
\]
If stage independence is established structurally for a control simulator, one may instead set $\rho=\tau=0$ exactly.
This is a consequence of the control SCM, not a conclusion from a small empirical dependence estimate.

For $K$ prespecified simulator settings, use $\eta=\gamma/K$ in each setting.
Another union bound provides simultaneous coverage of all $K$ population harms at least $1-\gamma$, without requiring independent simulations across settings.
The audit uses $K=120$ and $\gamma=0.01$.
The guarantee is a consequence of the sampling and projection argument; observing containment of empirical harms is not a test of population coverage.

\paragraph{Numerical dual repair and its scope.}
For a minimization LP $\min c^\top x$ with $Ax\le b$, $\mathbf1^\top x=1$, $x\ge0$, a dual pair satisfies $y\le0$, $A^\top y+z\mathbf1\le c$.
After replacing any positive numerical entries of $y$ by zero, set
$v=\max\{0,\max_j(A^\top y+z\mathbf1-c)_j\}$.
In exact arithmetic, $(y,z-v)$ is dual feasible, and $b^\top y+z-v$ is a valid lower bound.
Negating a maximization objective produces an upper bound in the same way.
The implementation adds further numerical padding and reports solver tolerances.
Finite precision in linear algebra and binomial quantiles remains a numerical consideration; this elementary repair does not establish a formally verified real-arithmetic computation.

\section{Scope and provenance of the extension arguments}
\label{sec:ext-scope}
The constructions use established tools: Fr\'echet couplings, total-variation event bounds, maximal couplings, one-dimensional transport, finite-class Hoeffding concentration, and binary-chain response composition.
Those tools and the existence of dynamic counterfactual compatibility restrictions are not claimed as new.
The dedicated results requiring comparison with nearest prior work are the calibrated joint bounds in the present shared-state model, the explicit sharp two-budget surface, and their connection to the exact three-state entrance.
General unknown-product TV optimization, arbitrary multivalued multistage sharpness, and data identification of the sensitivity budgets are not solved here.

\section{Small-state projection of simultaneous primitive confidence boxes}
\label{sec:ternary-projection-proof}
This section concerns the original independent-stage shared-response model.
It does not by itself permit stage dependence or direct response mismatch.
Confidence-region propagation and binomial tail inversion are established tools;
the specialization here is an eight-cell projection that is exact when the primitive boxes collapse to compatible points.

\paragraph{A second exact centered representation.}
For $n=3$, put $d_i=|\alpha_i-\beta_i|/2$ and
$b_i=\min\{(\alpha_i+\beta_i)/2,1-(\alpha_i+\beta_i)/2\}$.
For a binary vector $z$, let $c_i(z)=1$ if its nontrivial cut is the singleton $\{i\}$ or its complement, and zero otherwise.
Thus all $c_i$ vanish on the constant vectors, and exactly one is one on every nonconstant vector.
If $E_V=V-(\alpha-\beta)^\top p/2$, then
\begin{equation}
 E_L=\min_{Q:\mathbb E_Q Z=p}\sum_z Q_z\sum_ic_i(z)d_i,
 \qquad
 E_U=\max_{Q:\mathbb E_Q Z=p}\sum_z Q_z\sum_ic_i(z)b_i.
 \label{eq:projection-centered}
\end{equation}
The upper representation is the three-state cut theorem.
For the lower representation, start with $C_{ii}=\min(\alpha_i,\beta_i)$.
The remaining row and column masses have disjoint supports: each state has either row surplus or column deficit, never both.
Any transport between these remaining masses completes a coupling while preserving all diagonal maxima.
For every singleton cut it then has crossing probability
$\alpha_i+\beta_i-2C_{ii}=|\alpha_i-\beta_i|$.
Every three-state cut is a singleton, its complement, or a trivial cut.
This one coupling therefore minimizes crossing for all response cuts simultaneously, proving the lower representation and its attainment.
For $n<3$, append zero-mass states with known response zero; this changes no original potential outcome or endpoint.

\paragraph{Box coefficients and response projection.}
Suppose the true primitives belong to boxes
$\alpha_i\in[\alpha_i^-,\alpha_i^+]$, $\beta_i\in[\beta_i^-,\beta_i^+]$,
$p_i\in[p_i^-,p_i^+]$, and $m_a\in[m_a^-,m_a^+]$ for $a=A,B$.
The marginal state-law boxes may first be tightened by intersecting them with their probability simplices.
The exact coordinate projections of a nonempty simplex--box intersection are
\[
 \widetilde\alpha_i^-=
 \max\{\alpha_i^-,1-\sum_{j\ne i}\alpha_j^+\},\qquad
 \widetilde\alpha_i^+=
 \min\{\alpha_i^+,1-\sum_{j\ne i}\alpha_j^-\},
\]
and analogously for $\beta$.
Replace the original boxes by these tightened boxes below.
This retains every feasible state law; it does not assume the coordinate choices are jointly unconstrained.
Define
\begin{align}
 d_i^-&=\tfrac12\max\{0,\alpha_i^- -\beta_i^+,\beta_i^- -\alpha_i^+\},\\
 b_i^+&=\min\left\{\tfrac12,
             \tfrac12(\alpha_i^++\beta_i^+),
             1-\tfrac12(\alpha_i^-+\beta_i^-)\right\}.
 \label{eq:projection-coefficients}
\end{align}
Uniformly over the boxes, $d_i^-\le d_i$ and $b_i\le b_i^+$.
The first inequality is the distance between two intervals.
For the second, if $w_i\in[w_i^-,w_i^+]$, the maximum of
$\min(w_i,1-w_i)$ on that interval is
$\min(1/2,w_i^+,1-w_i^-)$.

Let
\[
 \mathcal Q_{\rm box}=\left\{Q\ge0:
 \sum_z Q_z=1,\quad p_i^-\le\sum_zQ_zz_i\le p_i^+\text{ for each }i\right\}.
\]
Compute the two eight-variable linear programs
\[
 e^- =\min_{Q\in\mathcal Q_{\rm box}}\sum_zQ_z\sum_ic_i(z)d_i^- ,
 \qquad
 e^+ =\max_{Q\in\mathcal Q_{\rm box}}\sum_zQ_z\sum_ic_i(z)b_i^+ .
\]
For every primitive triple in the boxes, its response polytope is contained in
$\mathcal Q_{\rm box}$, and~\eqref{eq:projection-centered} gives
$e^-\le E_L$ and $E_U\le e^+$.
Set $\Delta_-=m_A^- -m_B^+$ and $\Delta_+=m_A^+-m_B^-$.
Consequently the entire true identified interval is contained in
\begin{equation}
 \left[
 \max\{0,\Delta_-,e^-+\Delta_-/2\},\quad
 \min\{m_A^+,1-m_B^-,e^++\Delta_+/2\}
 \right].
 \label{eq:projection-final}
\end{equation}
This projection is generally conservative for nonzero boxes: coefficient extrema and endpoint-mean extrema need not be jointly attainable.
If all boxes collapse to points compatible with the original model,
the two LPs reduce to~\eqref{eq:projection-centered}; hence~\eqref{eq:projection-final}
equals the exact interval $[L,U]$ in exact arithmetic.

\paragraph{Sampling guarantee and incompatible boxes.}
For each of the $n$ coordinates of $\alpha$, $\beta$, and $p$, and for both endpoint means,
construct a two-sided binomial interval by tail inversion with failure allocation
$\zeta/(3n+2)$.
State counts have binomial coordinate marginals under iid arm sampling; response risks require their declared state-intervention sampling design.
Within each relevant batch, the Bernoulli sampling model must hold, conditional on sample size if that size is random.
The $3n+2$ confidence events need not be independent.
Their union bound implies that~\eqref{eq:projection-final} contains the entire true identified range with probability at least $1-\zeta$.
An unobserved state-intervention risk uses $[0,1]$ and contributes no failure event.
Zero-mass states introduced only for padding have both state laws and risks fixed to zero, so no extra error allocation is required.

The procedure must be defined even off the simultaneous coverage event.
If otherwise valid probability boxes are incompatible with the state-law simplices, or if their projected lower endpoint exceeds their upper endpoint, return $[0,1]$.
This vacuous fallback preserves the unconditional coverage guarantee.
Malformed inputs, such as negative counts or reversed primitive intervals, remain input errors.
Failure of a numerical optimizer should not be converted into an unsupported informative bound.
The proof is an exact-arithmetic statement; the supplied computation reports the separate effects of binomial-quantile arithmetic and repaired numerical LP dual bounds.

\section{Confidence projection and equal-observation-budget evaluation}\label{sec:budget-experiments}
The CP response projection uses the same six configurations and all 24,000 saved randomized datasets as the retained Lipschitz/Hoeffding procedure. Its simultaneous allocation is fixed at $\zeta/(3n+2)$ per probability coordinate, with no tuning of that allocation against the evaluation outcomes. These are method-development simulations reusing a fixed design, not preregistered or held-out performance estimates. It uses two eight-cell response LPs. Point-box and general-box tests are separate from the simulation coverage experiment. The new method has smaller mean width than the Lipschitz/Hoeffding procedure in all 24 strata, by $.00457$--$.11633$; this empirical comparison does not establish uniform dominance over all data or parameter values. All 24,000 CP-projection intervals contain their true structured identified ranges, again demonstrating conservative performance rather than exact nominal calibration. No two separately valid 95\% methods are intersected without a joint error allocation.

\subsection{Comparable observation counts and stronger endpoint baselines}
The structured methods use $N$ patients in each randomized arm, including state and endpoint, plus $N$ state-intervention outcomes for each of three states: total $5N$. To compare at the same observation count, each endpoint-only arm retains its original $N$ outcomes and adds $3N/2$ independent Bernoulli outcomes at that arm's true marginal risk. Thus it receives $5N/2$ outcomes per arm and $5N$ in total. All evaluated $N$ are even. The added draws use a fixed seed, 202609073033, once; every endpoint baseline uses the same combined counts. True parameters are used to generate the samples, never as inputs to the confidence algorithms. Equal observation counts do not imply equal monetary, clinical, or ethical costs of the interventions.

Three fixed endpoint procedures are retained. CP rectangle uses two-sided Clopper--Pearson intervals with failure $\zeta/2$ per arm and projects them through Fr\'echet bounds. Hoeffding rectangle uses the analogous two-sided concentration intervals with the same allocation. The stronger three-event procedure gives one-sided exact CP bounds $u_A$ on $m_A$ and $\ell_B$ on $m_B$, each with failure $\gamma=\zeta/3$, and sets
\[
 d_L=\widehat m_A-\widehat m_B-
 \sqrt{\tfrac12\log(1/\gamma)(1/n_A+1/n_B)}.
\]
Independent observations within and across arms imply the weighted Hoeffding bound for $d_L\le m_A-m_B$. A union bound over these three events then gives at least $1-\zeta$ coverage of the complete endpoint-only Fr\'echet range by
$[\max(0,d_L),\min(u_A,1-\ell_B)]$.
Exact binomial enumeration checks 147 small arm-size/risk combinations, including unequal arm sizes and boundary risks; it corroborates rather than replaces this proof.

The common comparison target is the true structured identified range. The endpoint procedures additionally cover their own potentially larger endpoint-only range, and both coverage targets are reported. The three-event procedure's empirical coverage ranges from $.997$ to $1$ for the structured range and $.964$ to $.991$ for its own endpoint-only range. None of these are claims of exact nominal calibration. A counterfactual coupling is not supplied as point-valued harm truth.

\subsection{Complete width comparisons}
The CP projection is narrower on average than either CP rectangle or the three-event endpoint procedure in 12 of 24 strata, and wider in 12. At $N=100$, all six configurations favor endpoint-only inference; at $N=10000$, the CP projection beats the three-event procedure in five of six settings, with mean reductions $.10287$--$.17841$. Setting 6 remains $.00571$ wider. The three-event procedure has smaller mean width than CP rectangle in all 24 evaluated strata, so it is retained rather than discarded in favor of a weaker baseline. Original structured intervals beat equal-budget CP in only 8 of 24 strata. These results expose the finite-sample cost of additional unknown primitives while showing how their structural information helps at larger sample sizes. The fixed settings and all failures, widths and paired Monte Carlo errors are retained in the package.

\begin{table}[H]\centering\caption{Mean interval widths for all configurations. Every procedure uses $5N$ observations. LH: retained Lipschitz/Hoeffding; projection: proposed CP response LP; CP/H: endpoint rectangles; 3-event: stronger endpoint comparator.}\begin{tabular}{crrrrrr}\toprule Setting & $N$ & LH & Projection & CP & H & 3-event\\\midrule
1 & 100 & 0.6353 & 0.6166 & 0.5535 & 0.5746 & 0.5500\\
1 & 500 & 0.5600 & 0.4757 & 0.5243 & 0.5341 & 0.5227\\
1 & 2000 & 0.4583 & 0.4040 & 0.5121 & 0.5170 & 0.5112\\
1 & 10000 & 0.3890 & 0.3644 & 0.5053 & 0.5075 & 0.5049\\
2 & 100 & 0.7600 & 0.6818 & 0.5044 & 0.5766 & 0.4882\\
2 & 500 & 0.5149 & 0.3986 & 0.3894 & 0.4233 & 0.3839\\
2 & 2000 & 0.3521 & 0.2782 & 0.3424 & 0.3598 & 0.3401\\
2 & 10000 & 0.2475 & 0.2140 & 0.3178 & 0.3256 & 0.3168\\
3 & 100 & 0.6515 & 0.6273 & 0.5616 & 0.5848 & 0.5557\\
3 & 500 & 0.5641 & 0.5011 & 0.4975 & 0.5231 & 0.4898\\
3 & 2000 & 0.4666 & 0.3858 & 0.4503 & 0.4655 & 0.4465\\
3 & 10000 & 0.3613 & 0.3146 & 0.4237 & 0.4306 & 0.4220\\
4 & 100 & 0.6804 & 0.6575 & 0.5921 & 0.6186 & 0.5840\\
4 & 500 & 0.5888 & 0.5146 & 0.5129 & 0.5420 & 0.5046\\
4 & 2000 & 0.4479 & 0.3875 & 0.4653 & 0.4805 & 0.4614\\
4 & 10000 & 0.3421 & 0.3137 & 0.4387 & 0.4455 & 0.4370\\
5 & 100 & 0.5369 & 0.4926 & 0.4552 & 0.4767 & 0.4517\\
5 & 500 & 0.4351 & 0.3260 & 0.4238 & 0.4341 & 0.4222\\
5 & 2000 & 0.3172 & 0.2605 & 0.4118 & 0.4171 & 0.4111\\
5 & 10000 & 0.2523 & 0.2264 & 0.4052 & 0.4076 & 0.4048\\
6 & 100 & 0.4256 & 0.4032 & 0.3294 & 0.3556 & 0.3260\\
6 & 500 & 0.3366 & 0.3232 & 0.2922 & 0.3051 & 0.2907\\
6 & 2000 & 0.2991 & 0.2915 & 0.2768 & 0.2835 & 0.2761\\
6 & 10000 & 0.2789 & 0.2743 & 0.2689 & 0.2720 & 0.2686\\
\bottomrule\end{tabular}\end{table}
\section{Extended experiment protocols}\label{sec:new-experiments}
All extension inputs are generated by fixed seeds before comparison. No case is filtered on whether it supports the analytic bound. The code distinguishes mathematical formulas and rational certificates from floating-point optimization diagnostics.
\subsection{General mechanism and multistage tests}
The joint-sensitivity audit uses 30 cases each at $n=1,2,3,4,6$, Dirichlet-generated state/response laws, direct-response mixtures, and product-to-dependent coupling mixtures. Primitive risks, budget values, regime means, actual harm, and the outer interval are retained. This is containment for known generated mechanisms, not a claim of global sharpness. The independent direct-effect reference uses 12/24/6 primitive configurations at $n=2/3/4$ and three mismatch-budget scales, for 126 cases. Enumerating transport vertices and solving a joint two-response-vector LP at each is globally sufficient because, with independent stages, the response feasible set does not depend on the transport. General joint TV sensitivity does not have that separation.
\begin{figure}[t]
\centering
\includegraphics[width=.58\linewidth]{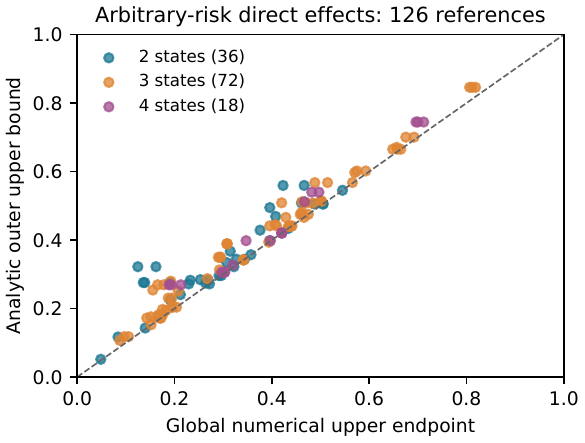}
\caption{General direct-effect sensitivity at arbitrary risks. Each point compares an analytic outer upper bound with a global numerical upper endpoint; 126 configurations cover two, three and four states. The diagonal marks equality. Positive gaps demonstrate that the general enclosure can be loose, unlike the symmetric frontier in Theorem~\ref{thm:surface}.}
\label{fig:direct-reference}
\end{figure}
The multistage audit uses eight cases at each binary suffix length $1,2,4,6$, retaining strictly interior transition risks. It composes every entrance vertex pair and every binary response-polytope vertex, then compares the minimum and maximum terminal harm with the product formula. The symbolic construction, rather than these finite checks, establishes arbitrary suffix length. Independent audits include 81 exact rational joint-frontier witnesses, 240 endpoint-stability comparisons using full small-model enumeration, 49 additional independently constructed rational witnesses, 80 two-state direct-effect checks, 2,000 centered-stability comparisons, and 3,600 joint-sensitivity sampling checks. Separate audit scripts and results retain their scopes.

\subsection{Retained Lipschitz/Hoeffding baseline and information comparison}
Six settings estimate all state and response probabilities. Each repetition independently generates arm state counts and their state-conditional Bernoulli outcomes, plus three separate state-intervention outcome counts. Thus the same two arm samples estimate both state laws and endpoint means. Dependence between these confidence events is allowed by the union bound. All $M_i=N>0$ in these experiments. For $M_i=0$, the interface sets $\widehat p_i=\epsilon_i=1/2$, giving the vacuous risk interval $[0,1]$; it rejects invalid or noninteger counts.

The endpoint-only comparator allocates failure $.025$ to each arm's Hoeffding interval. If these are $[\ell_A,u_A]$ and $[\ell_B,u_B]$, its confidence outer interval is $[\max(0,\ell_A-u_B),\min(u_A,1-\ell_B)]$. It uses the same $2N$ arm outcomes; the Lipschitz/Hoeffding baseline additionally uses $3N$ state-intervention outcomes. Each method separately has at least 95\% coverage. Their intersection does not automatically retain 95\% coverage without a joint failure allocation. Across 24 strata, the Lipschitz/Hoeffding procedure has smaller mean width in 12 and larger mean width in 12. The largest mean reduction is $.15959$ and the largest increase is $.03423$. At $N\ge2000$ it is narrower in five of six settings. Both methods contain the full true range in all 24,000 repetitions; this documents conservatism, not exact nominal calibration. A plug-in identified interval is also retained as a diagnostic, without a nominal confidence guarantee.

\begin{table}[H]\centering\caption{Unknown-primitive configurations and true sharp interval. Setting labels are one-based in the paper and zero-based in JSON.}\begin{tabular}{cllll}\toprule Setting & $\alpha$ & $\beta$ & $p$ & $[L,U]$\\\midrule
1 & $(1/3,1/3,1/3)$ & $(1/3,1/3,1/3)$ & $(0.5,0.5,0.5)$ & $[0.0000,0.3333]$\\
2 & $(0.1,0.35,0.55)$ & $(0.55,0.35,0.1)$ & $(0.1,0.45,0.85)$ & $[0.3375,0.5000]$\\
3 & $(0.15,0.25,0.6)$ & $(0.25,0.45,0.3)$ & $(0.8,0.05,0.6)$ & $[0.1100,0.3675]$\\
4 & $(0.02,0.48,0.5)$ & $(0.01,0.29,0.7)$ & $(0.02,0.9,0.3)$ & $[0.1140,0.3684]$\\
5 & $(0.5,0.2,0.3)$ & $(0.1,0.6,0.3)$ & $(0,0.5,1)$ & $[0.0000,0.2000]$\\
6 & $(0.7,0.15,0.15)$ & $(0.2,0.4,0.4)$ & $(0.15,0.65,0.4)$ & $[0.0000,0.2425]$\\
\bottomrule\end{tabular}\end{table}
\begin{table}[H]\centering\caption{Mean confidence-interval widths (Lipschitz--Hoeffding / endpoint-only), 1,000 repetitions per cell. Both have 100\% empirical set containment here; neither identifies actual harm.}\begin{tabular}{crrrr}\toprule Setting & $N=100$ & $N=500$ & $N=2000$ & $N=10000$\\\midrule
1 & 0.635 / 0.621 & 0.560 / 0.553 & 0.458 / 0.527 & 0.389 / 0.512\\
2 & 0.760 / 0.726 & 0.515 / 0.495 & 0.352 / 0.396 & 0.248 / 0.342\\
3 & 0.652 / 0.637 & 0.564 / 0.557 & 0.467 / 0.501 & 0.361 / 0.447\\
4 & 0.680 / 0.666 & 0.589 / 0.589 & 0.448 / 0.517 & 0.342 / 0.462\\
5 & 0.537 / 0.522 & 0.435 / 0.454 & 0.317 / 0.427 & 0.252 / 0.412\\
6 & 0.426 / 0.411 & 0.337 / 0.330 & 0.299 / 0.296 & 0.279 / 0.277\\
\bottomrule\end{tabular}\end{table}
The new CP projection and equal-observation-budget comparison are reported separately in Appendix~\ref{sec:budget-experiments}. These retained results describe the older confidence wrapper, not the proposed CP projection.
\subsection{Native simulator audits and applicability limits}
We retain the original EpiCare transition, remission, symptom, and adverse-event
mechanisms, including the default adverse-event threshold $0.999$. Across
$30$ seeds, $3/4$ diseases, and persistent patient modifiers on/off, we compare
the first lexicographic treatment pair with zero immediate remission, followed
by common state-adaptive care. The endpoint is an adverse event by the second
visit; first-visit events remain as an absorbing state, yielding $4/5$ states.
We specify a source-compatible cross-world SCM using shared within-visit ranks
and Gaussian innovations, independent visit shocks, and persistent modifiers.
Its coupling is an audit convention, not identified by intervention kernels.

Each of the $120$ cases uses $2^{20}$ i.i.d. simulated patients. Integer joint
tables are retained. Computation takes $179.4$ seconds on a local CPU. All
empirical-table sensitivity certificates hold. With persistent modifiers, mean
estimated dependence is $2.37\times10^{-5}$ and $4.02\times10^{-5}$ for $3/4$
diseases; the maximum is $1.87\times10^{-4}$. With modifiers disabled population
dependence is exactly zero, despite nonzero empirical estimates. Native rare
events yield no large, well-resolved violation of the independence interval;
we report this negative result rather than alter the simulator to amplify it.

For the fixed response $z_0$ with no terminal AE in active states, insertion of
$C\otimes\delta_{z_0}$ gives $\rho\le2\Pr(Z\ne z_0)$. Binomial upper confidence
bounds therefore calibrate conservative dependence budgets. We additionally
propagate simultaneous binomial confidence boxes for all state probabilities,
state-intervention risks, and arm means through transport LPs and the mean-shift
correction. Error allocation across all cases gives conservative joint $99\%$
coverage; these are outer confidence regions, not sharp intervals. Original
\texttt{EpiCare.step()} runs provide $80{,}000$ independent single-world episode
checks; all eight SCM event estimates lie in the corresponding native $99\%$
binomial intervals.

A second native-source audit uses the Oberst--Sontag sepsis simulator, with
$60$ prespecified initial-state/diabetes configurations and $240{,}000$ episodes.
The compared first-visit treatments retain different treatment switches in the
complete intermediate state; consequently all active supports are disjoint.
Common absorbers have deterministic outcomes. Positive thresholds on A-only
states and reversed thresholds on B-only states simultaneously attain every
pairwise upper cost, so this protocol admits no coherence gain over pairwise
transport. Deleting the treatment switches to create overlap would require a
new causal sufficiency argument. This second-source negative audit delineates
applicability; it is not real-patient validation or a second positive benchmark.

\paragraph{Native-source provenance and inference details.}
The EpiCare audit retains source revision \texttt{b4288ec5e43ca5fb6e9b6a5effad4d15aa507088} (Apache-2.0). The sepsis source is \url{https://github.com/clinicalml/gumbel-max-scm}, revision \texttt{d62995deb5cb51e3ecf2859f4cbbc13f651131a2} (MIT). Both snapshots and hash manifests are included. No credentials, human patient records, or remote training service is required. The native EpiCare configuration retains two symptom dimensions, four treatments, and the default threshold; the lower disease count is explicitly selected for the finite-state audit. The baseline fixed response for the rare-event dependence bound is $z_0=(0,\ldots,0,1)$, including the adverse-event absorber. Native persistent modifiers may couple the state pair and this common response vector without introducing a treatment-specific terminal mechanism. The all-primitives confidence boxes allocate failure $.01/120$ per case and divide that allocation across $3n+3$ binomial events, where $n$ counts the absorber. Simultaneous coverage is therefore at least 99\% over the complete set of 120 specified population mechanisms, under exact arithmetic. The implementation repairs LP dual inequalities and adds numerical margins; it is not a formal interval-arithmetic proof of SciPy's special functions. Appendix~\ref{sec:ext-simulator-confidence} proves the population bound. Containment of all empirical tables is a separate deterministic implementation check, not a repeated-sampling estimate of population coverage. An optional coarsened pointwise diagnostic gives small positive lower bounds in 3 of 60 persistent-modifier cases; this exploratory comparison is not multiplicity-adjusted and does not establish a large population effect.

For sepsis, diabetes is fixed within each evaluated configuration, and the complete Markov state includes physiology and active treatment switches. Native endpoint mortality is recorded for all 60 configurations. The support-separation claim follows from the source state updates; the sampled support overlap of zero corroborates it and is not used as proof of a population zero. For A-only state $i$ use $Z_i=\mathbf1\{V\le p_i\}$ and for B-only state $j$ use $Z_j=\mathbf1\{V>1-p_j\}$. Then $\Pr(Z_i=1,Z_j=0)=\min(p_i,1-p_j)$ simultaneously. Constant absorbing-state responses satisfy their cross-pair costs too. This proves pairwise attainability for arbitrary risk values in this protocol, while allowing pairwise transport itself to improve upon endpoint-only information.

\section{An arbitrary-state formula and certified computation}
\subsection{Balanced risks identify a partition problem}
\begin{theorem}[Pooled-law reduction and worst relaxation gap]\label{thm:pooled}
If $p_i=1/2$ for every state, then $L=0$ and
\begin{equation}
 U=b(w):=\max_{S\subseteq[n]}\min\{w(S),1-w(S)\},\qquad
 U_{\rm pair}=\min\{1/2,1-\max_iw_i\}.\label{eq:pooled}
\end{equation}
For all finite state spaces, $0\leq U_{\rm pair}-U\leq1/6$, and $1/6$ is attained by uniform three-state $w$.
\end{theorem}
For the upper bound, condition on a response cut and apply its maximal crossing probability in~\eqref{eq:disagree}. For attainability, choose a maximizing subset $S$, use a transport maximizing crossing of that cut, and give equal probability to the complementary response tables $\one_S$ and $\one-\one_S$. Every terminal marginal is $1/2$ and both tables use the same transport. The pooled law determines the answer even when the two treatment-specific state laws differ.

If a pooled atom is at least $1/2$, its complement attains $U=1-\max_iw_i$ and the pairwise bound is exact. Otherwise some subset has mass in $[1/3,2/3]$, so $U\geq1/3$ while $U_{\rm pair}=1/2$. This proves the worst-gap statement. Complementary-Bernoulli and balanced-subset constructions have precedents in dependence modeling and experimental design \citep{Huber_2019,morrison2025optimalcorrelationbernoullitrials}; the causal transport reduction and quantified sharp endpoints are the results here.

For uniform $\alpha=\beta$ and a common terminal risk $p$, the formula extends to
\begin{equation}
 U=\min\{p,1-p,\lfloor n/2\rfloor/n\}.\label{eq:uniform}
\end{equation}
A permutation coupling decomposes into cycles; each deterministic binary response has at most $\lfloor\ell/2\rfloor$ directed $1\to0$ transitions on a cycle of length $\ell$. A cyclic response-pattern construction attains the bound. The gap is therefore not confined to the single risk value $1/2$.

\subsection{Exact and approximate algorithms}
When pooled weights are supplied as nonnegative integers $a_i$ with $W=\sum_i a_i>0$ and $w_i=a_i/W$, standard subset-sum dynamic programming finds the largest achievable integer sum at most $W/2$; dividing by $W$ gives $U$ and the corresponding subset gives an attaining SCM. Under this integer-weight encoding, deciding whether $U=1/2$ is weakly NP-complete by PARTITION, even in this two-stage, binary-outcome, known-transition subclass. This is a restricted-model reduction; general multimarginal hardness already has precedents \citep{altschuler2020hardnessresultsmultimarginaloptimal}.

For additive error $\eta>0$, round each weight downward in units $\eta/n$, solve the rounded subset-sum problem with capacity $1/2$, and evaluate the chosen subset using original weights. If its smaller group mass is $\ell$, then
\begin{equation}
 \ell\leq U\leq\min\{U_{\rm pair},\ell+\eta\}.\label{eq:approx}
\end{equation}
There are $O(n^2/\eta)$ dynamic-program updates and $O(n/\eta)$ sum states; storing a witness mask costs $n$ bits per state in our implementation. If $\max_iw_i\geq1/2$, solve exactly. Otherwise $U\geq1/3$, so choosing $\eta=\varepsilon/3$ gives a multiplicative $(1-\varepsilon)$ approximation. The implementation interprets floating-point weights as exact binary rationals, normalizes them exactly, performs exact rounding, and rounds reported interval endpoints outward. The integer solver rejects noninteger weights. These numerical details matter for a certificate.

\section{Heterogeneous risk and sampling uncertainty}
The balanced-risk formula is not a license to replace arbitrary risks by $1/2$. A controlled extension follows from a continuity bound. For fixed $\alpha,\beta$ and two risk vectors $p,q$,
\begin{equation}
 \left|U(p)-U(q)-\tfrac12(\alpha-\beta)^T(p-q)\right|
 \leq\sum_iw_i|p_i-q_i|.\label{eq:continuity}
\end{equation}
To prove it, couple two response vectors by coordinate flips with probabilities $|p_i-q_i|$, independently of the first-stage mechanism. Their disagreement probabilities differ by at most $2\sum_iw_i|p_i-q_i|$. Equation~\eqref{eq:disagree} gives the result. Thus~\eqref{eq:pooled} provides a computable enclosure for the sharp upper endpoint at heterogeneous risks. This enclosure is generally not sharp; for three states one should instead use~\eqref{eq:three}.

For the balanced-risk subclass, $b(w)$ is one-Lipschitz in total variation. With $N$ independent randomized samples per arm, let $\widehat w=(\widehat\alpha+\widehat\beta)/2$. A union bound over subsets and Hoeffding's inequality give
\begin{equation}
 \Pr\left\{|b(\widehat w)-b(w)|\leq
 r_N\right\}\geq1-\zeta,\qquad
 r_N=\sqrt{\frac{\log(2^n/\zeta)}{4N}}.\label{eq:stat}
\end{equation}
Intersect the computed endpoint interval enlarged by $r_N$ with $[0,1/2]$. This is a confidence interval for the \emph{sharp upper endpoint}, conditional on known $p_i=1/2$ and the specified causal model. It is not a confidence interval identifying the actual individual harm, nor a solution to unmeasured confounding. Unknown terminal risks can be enclosed separately using~\eqref{eq:continuity}; this older diagnostic is conditional on known terminal risks; Section~\ref{sec:finite-sample} and Appendix~\ref{sec:ext-finite-sample} provide the separate all-primitives extension.

\section{Additional baseline validation and declared mechanism stress}\label{sec:legacy-evaluation}
We test the algebraic claims against exhaustive compatible mechanisms and then examine a public medical decision benchmark. All experiments use local CPU computation and publicly available code or generated probabilities. No training of a foundation model is needed to evaluate the sharpness claims.

\paragraph{Exhaustive reference and independent checks.}
For $n\leq4$, we enumerate every feasible vertex of the first-stage transport polytope and every feasible vertex of the $2^n$-cell response polytope, then evaluate all pairs. The bilinear objective attains each global extremum at such a vertex pair. This is an exhaustive specialization of joint-response optimization, not an alternating local solver. The uniform-risk upper formula agrees in 27 configurations. Ninety random configurations verify the exact balanced-risk formula and, separately, containment by the heterogeneous-risk enclosure; no general heterogeneous four-state closed form is tested or claimed. An additional 600 random configurations compare the cut envelope with global enumeration: it agrees up to $2.22\times10^{-16}$ for $n\leq3$ and can be strict for $n=4$. The three-state closed form is separately checked against its response LP on 3,000 configurations; an independently written check adds 1,500 configurations and joint-cell feasibility tests. No random instance is omitted because its gap vanishes.

\paragraph{Public medical primitives and explicit adaptations.}
EpiCare is a public, medically inspired dynamic-treatment benchmark \citep{NEURIPS2024_eb868633}. We pin its source revision and extract two-visit subproblems with three or four latent disease states and four treatments. The full disease state is exposed for an oracle mechanism audit. We disable persistent patient modifiers to respect independent stage noises and disable adverse-event termination; the binary endpoint is failure to remit on the final visit. Initial actions are chosen to have zero immediate remission, avoiding conditioning on survival. The common terminal care is either fixed action $0$ or a state-dependent remission-maximizing rule used identically in both worlds. These configurations are not a validated clinical trial.

For each $n\in\{3,4\}$ and seed $0,\ldots,29$, we retain the first eligible initial state/action pair by a deterministic rule. Two efficacy ranges are specified through EpiCare's public parameter interface: its native $[0.8,1]$ and an intermediate-risk stress range $[0.35,0.65]$. To test transition diffuseness, we additionally define
\begin{equation}
 T_\lambda=(1-\lambda)T+\lambda\one\one^T/n,
 \qquad\lambda\in\{0,1/2,1\},\label{eq:stress}
\end{equation}
before applying the original treatment transition modifiers and row normalization. The $\lambda>0$ cases are explicitly \emph{EpiCare-derived mechanism stress tests}, not unchanged EpiCare or a clinically calibrated alternative. All 720 prespecified cases are retained; the accompanying data also retain all 1,916 eligible upstream-transition cases and source primitives.

\paragraph{Baselines and observed boundary.}
We compare the sharp interval with endpoint-only Fr\'echet bounds, pairwise transport~\eqref{eq:pair}, and the pure-mediator summand upper bound of \citet{ijcai2022p376}, also present in the multivalued extension of \citet{shu2026generalprobabilitiescausationcausal}. The latter is
\[
 U_{\rm med}=\min\left\{m_A,1-m_B,
 \sum_{i\ne j}\min(p_i,1-p_j)\min(\alpha_i,\beta_j)\right\}.
\]
It is a valid non-sharp comparator, not an incorrectly advertised sharp method.

All 240 cases with upstream transitions ($\lambda=0$, including both efficacy ranges) have zero pairwise-to-sharp gap. All 240 have maximum pooled-state mass at least $0.64599$, so Proposition~\ref{prop:dominant} explains exactness even at heterogeneous terminal risks. Positive gaps appear only in the declared mixing and intermediate-efficacy stress settings with state-adaptive common care. Table~\ref{tab:medical} reports every seed in that stratum; other strata remain zero. This negative baseline is part of the finding: an evaluation using only the native transition mechanism would not exercise this compatibility gap.

\begin{table}[t]
\caption{EpiCare-derived intermediate-efficacy, state-adaptive-care stress stratum, 30 seeds in every row. Gap is pairwise upper minus sharp upper, in probability units. All other care/efficacy strata have zero pairwise gap.}
\label{tab:medical}\centering
\begin{tabular}{ccrrrr}
\toprule
States & Mixing $\lambda$ & Positive gaps & Mean gap & Maximum gap & Max. cut gap\\
\midrule
3 & 0 & 0/30 & 0 & 0 & 0\\
3 & $1/2$ & 2/30 & .00157 & .02466 & 0\\
3 & 1 & 30/30 & .06828 & .11347 & 0\\
4 & 0 & 0/30 & 0 & 0 & 0\\
4 & $1/2$ & 9/30 & .00605 & .04682 & .00995\\
4 & 1 & 30/30 & .01665 & .05270 & .01993\\
\bottomrule
\end{tabular}
\end{table}

\begin{figure}[t]
\centering\includegraphics[width=\linewidth]{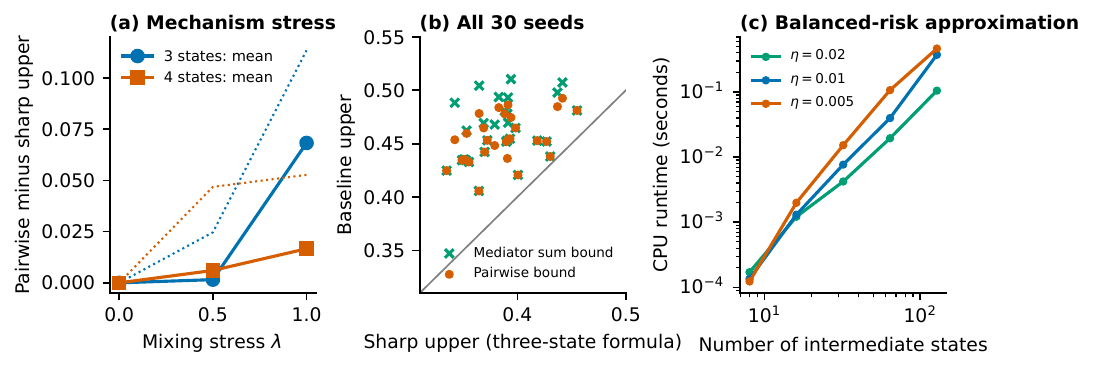}
\caption{(a) Mean (solid) and maximum (dotted) pairwise gap across the 30-seed intermediate-efficacy, adaptive-care stratum. (b) Every three-state, $\lambda=1$ seed compared with two known outer bounds. (c) Local runtime of the certified balanced-risk approximation; it does not solve the unrestricted heterogeneous-risk problem. Native-transition and other zero-gap strata are retained in the data and described in the text.}
\label{fig:experiments}
\end{figure}

\paragraph{Numerics, finite samples and scale.}
The 720 global small-model bounds take 3.35 seconds locally after source extraction, with constraint residuals below $5.56\times10^{-16}$. Source-kernel validation uses 5,000 independent simulator draws in each of 16 configurations; the largest absolute frequency errors are $0.01186$ for transitions and $0.00683$ for remission. These checks verify implementation, not clinical validity. The balanced approximation is evaluated up to 128 states for three error tolerances. A separate finite-sample experiment uses four known balanced-risk settings, 1,000 repetitions and $N\in\{100,500,2000,10000\}$ per arm, checking coverage of~\eqref{eq:stat}. The supplement reports the protocol and coverage; the package retains exact seeds, measured runtimes, per-case bounds and all finite-sample repetitions. Probability-bound comparisons are deterministic within each generated mechanism; we do not treat simulator seeds as an iid clinical cohort or report clinical significance tests.

\section{Complete experiment protocol and reproduction}
\label{sec:experiments-supp}
The executable research package contains a single local reproduction entry point, the original probability inputs, generated JSON summaries, figures, source licenses and provenance, and independent mathematical checks. No account or patient-data application is required. Reading a simulator's specified primitive probabilities is used only as an oracle experiment; their availability in an actual clinical dataset is not assumed.

\subsection{Exhaustive optimization reference}
For $n\leq4$, the transport equality matrix contains all $n$ row constraints and the first $n-1$ column constraints. Every subset of $2n-1$ variables is checked for a full-rank basis. Solving that basis and keeping nonnegative solutions enumerates all vertices; degenerate duplicates are removed. The response equality matrix has a normalization row and the $n$ binary marginal rows over all $2^n$ response patterns. Every subset of $n+1$ variables is similarly checked. The bilinear objective is evaluated for every pair of vertices. For each fixed argument a linear objective reaches its extremum at a vertex of the other polytope; selecting vertices successively proves that a global vertex-pair optimum exists.

The generic reference uses floating-point linear algebra, a $10^{-10}$ feasibility allowance, and a maximum final equality residual check of $10^{-8}$. Deduplication rounds keys to 11 decimal places but retains unrounded solutions. Thus the generic results are numerical checks, not formal exact rational certificates. The supplied four-state counterexample additionally uses exhaustive integer determinants and rational arithmetic without rank or feasibility tolerances; Appendix~\ref{sec:supp-four} records its certificate. Our closed-form and exact integer-subset algorithms are separate from the numerical reference.

For 27 homogeneous configurations, $n\in\{2,3,4\}$ and $p\in\{.1,.25,1/3,.4,.5,.6,2/3,.75,.9\}$. The 90-case validation uses 30 cases per $n$, independently sampled Dirichlet$(\one)$ first-stage laws and Uniform$[.05,.95]$ heterogeneous risks, with seed 270907. It checks the pooled-risk formula, attaining SCM witness, rectangular-gap formula, heterogeneous-risk enclosure and global reference. A separate seed 280907 generates 200 cases per state count with Uniform$[0,1]$ risks, comparing the cut envelope to global enumeration. For the three-state closed form, seed 290907 generates 1,000 cases each for $n=1,2,3$ and checks the response LP; zero-mass padding is included. An independent script uses its own construction and verifies 1,500 three-state cases, including the feasibility of the returned eight-cell response distribution. The integer subset-sum routine is checked on 120 integer-weight instances against exhaustive subset enumeration. Two additional independently written dominant-state checks use 240 and 120 configurations, including boundary masses and terminal risks; both compare the star construction with a transport LP. The second constructs the complete threshold-segment response distribution. Across all 240 upstream-transition simulator cases, the maximum pooled-state mass ranges from $0.64599$ to $1$, verifying the sufficient condition of Proposition~\ref{prop:dominant}.

\subsection{Public EpiCare extraction}
The repository is \url{https://github.com/Grosenick-Lab-Cornell/EpiCare}, pinned at commit \texttt{b4288ec5e43ca5fb6e9b6a5effad4d15aa507088}. The unmodified environment source and Apache-2.0 license are included with the minimal reproduction files. Its source SHA-256 is recorded in the provenance ledger. The source was accessed on September 7, 2026.

The shared configuration is four treatments, two symptom dimensions, two visits, \texttt{patient\_modifiers=False}, and \texttt{adverse\_event\_threshold=1.0}. The last choice disables toxicity stopping because the expit symptom values are at most one and stopping uses a strict greater-than-threshold test. This intentionally isolates remission and is not evidence about adverse-event safety. For every seed and state count, the first eligible initial state and pair of distinct initial treatments (enumerated as combinations and oriented with the smaller action index first) with zero remission probability are selected lexicographically. No case is selected on its counterfactual gap. The full list of eligible choices is also retained.

In the original step order, remission is checked before drawing a new disease state. Selecting initial actions with zero remission therefore gives unconditional first-stage disease laws. For initial disease $s$ and treatment $a$,
\[
 \mu_a(j)=\frac{T(s,j)m_a(j)}{\sum_kT(s,k)m_a(k)},
\]
where $m_a$ is the original treatment transition modifier. Terminal remission risk at disease $j$ is read from the original remission table for the fixed common care rule. The adverse outcome in our experiment is its complement, failure to remit. In the adaptive-care stratum, the same rule chooses the remission-maximizing action at each disease state in both worlds. Full latent state exposure makes this an oracle rule, not an implementable recommendation from noisy symptoms.

Efficacy ranges are $[.8,1]$ (native parameter range) and $[.35,.65]$ (explicit efficacy stress). Disease persistence is hard coded rather than exposed as a constructor parameter, so the additional mixing modification in the main text is applied transparently to the extracted base transition matrix. All treatment modifiers and remission tables remain those of the corresponding source-generated instance. The input records retain original and effective transition matrices, modifiers, initial choices, terminal care vectors, event probabilities, and whether each record is upstream-transition or derived stress. The 720 comparisons cross two state counts, 30 seeds, two efficacy settings, two care definitions and three mixing strengths. The $\lambda=0$ subset has 240 records. No improvement over the pairwise transport bound is claimed in that subset.

The source extraction validates the analytic primitives against 16 configurations with 5,000 simulator draws each. Maximum transition and remission frequency discrepancies are $0.01186$ and $0.00683$. Sampling variability is expected; the exact kernel formulas, source order, fixed seed and all discrepancies are recorded. Source extraction takes approximately 61.44 seconds locally, separate from the 3.35-second bound computation. The optional source-regeneration path uses Gym 0.23.1; reproducing the bound analysis from the included public primitive JSON requires only NumPy, SciPy and Matplotlib.

\clearpage
\subsection{Finite-sample audit and what its coverage means}
The four settings used for~\eqref{eq:stat} are
\begin{center}
\begin{tabular}{cll}
\toprule
Setting & $\alpha$ & $\beta$\\
\midrule
1 & $(1/3,1/3,1/3)$ & $(1/3,1/3,1/3)$\\
2 & $(.10,.35,.55)$ & $(.55,.35,.10)$\\
3 & $(.15,.25,.60)$ & $(.25,.45,.30)$\\
4 & $(.02,.48,.50)$ & $(.01,.29,.70)$\\
\bottomrule
\end{tabular}
\end{center}
\begin{table}[H]
\centering
\caption{Empirical coverage of the nominal 95\% endpoint interval over 1,000 repetitions. The union bound is conservative in these settings.}
\begin{tabular}{crrrr}
\toprule
Setting & $N=100$ & $N=500$ & $N=2000$ & $N=10000$\\
\midrule
1 & 1.000 & 0.999 & 0.999 & 0.998\\
2 & 1.000 & 0.998 & 1.000 & 0.998\\
3 & 1.000 & 1.000 & 0.998 & 1.000\\
4 & 0.998 & 0.999 & 1.000 & 0.999\\
\bottomrule
\end{tabular}
\end{table}

For every setting, terminal risks are known and exactly $1/2$. Each of 1,000 repetitions draws two independent multinomial count vectors with $N\in\{100,500,2000,10000\}$ samples per arm. Pooled counts have exact integer denominator $2N$ and are processed by the integer subset-sum solver. The Hoeffding radius, point error, coverage and width are retained for every setting and sample size. These are confidence intervals for the sharp upper endpoint, not for the unknown realized harm, and do not establish coverage for unknown heterogeneous terminal risks. The lower endpoint of the actual identified set is zero throughout this audit.

\end{document}